\documentclass[preprint2,10pt,a4paper]{aastex}
\usepackage{amsmath}
\usepackage{graphicx}
\usepackage{multicol}
\begin{document}
\title{Magnetar Oscillations I: strongly coupled dynamics of the crust and the core}
\author{Maarten van Hoven and Yuri Levin}
\affil{Leiden University, Leiden Observatory and Lorentz Institute,
 P. O. Box 9513, NL-2300 RA Leiden}
\email{vhoven@strw.leidenuniv.nl, yuri@strw.leidenuniv.nl}
\begin{abstract} \noindent
Quasi-Periodic Oscillations (QPOs) observed at the tail end of Soft Gamma Repeaters giant flares
are commonly interpreted
as the torsional oscillations of magnetars.
From a theoretical perspective, the oscillatory motion is influenced
by the strong
interaction between the shear modes of the crust and magnetohydrodynamic 
Alfven-like modes in the core.
We study the dynamics which arises through this interaction, and present
several new results:
(1) We show that discrete {\it edge modes} frequently reside near the
edges of the core Alfven continuum,
and explain using simple models why these are generic and long-lived.
(2) We compute the magnetar's oscillatory motion for realistic
axisymmetric magnetic field configurations and core density profiles,
but with a simplified model of the elastic crust. We show that one may generically get
multiple gaps in the Alfven continuum. One obtains strong discrete {\it gap modes}
if the crustal
frequencies belong to the gaps; the resulting frequencies do not coincide 
with, but are in some cases close to the crustal frequencies.
(3) We deal with the issue of tangled magnetic fields in the core by developing a phenomenological
model to quantify the tangling. We show that  field tangling enhances the role of the core discrete
Alfven modes and reduces the role of the core Alfven continuum in the overall oscillatory dynamics
of the magnetar.
(4) We demonstrate  that the system displays transient QPOs when  parts of 
the spectrum of
the core 
Alfven modes contain discrete modes which are densely and 
regularly spaced in frequency. The transient QPOs are the strongest when they
are located near the frequencies of the crustal modes.
(5) We show that if the neutrons are coupled into the core Alfven motion,
then 
 the post-flare crustal motion is strongly damped and has a very weak amplitude.
We thus argue that
magnetar QPOs give evidence that the proton and neutron components in the
core are dynamically decoupled and that at least
one of them is a quantum fluid. 
(6) We show that it is difficult to
identify the
 high-frequency 
$625$Hz QPO as being due to the physical oscillatory mode of the magnetar, if the latter's fluid core
consists of the standard proton-neutron-electron mixture and is magnetised to the same extent as the
crust.

\end{abstract}
\section{Introduction}
Since the discovery of quasi periodic oscillations (QPOs) in the
lightcurves of giant flares from soft gamma repeaters (SGR)
(Israel et al., 2005; Strohmayer \& Watts, 2005; Watts \& Strohmayer,
2006; Barat et al., 1983) there has been  considerable interest in their
physical origin.
One of the appealing explanations is that the QPOs are driven by torsional
oscillations\footnote{By torsional
oscillations we mean those which are nearly incompressible. Modes with
compression have high restoring
forces and feature much higher frequencies than most of the observed
QPOs.}  of the
neutron stars whose magnetic energy
powers the flares (Duncan 1998).
This opens a unique possibility to perform an asteroseismological analysis of
neutron stars, and possibly obtain a new observational window to study the
neutron-star
interiors. Many authors have considered torsional modes to be confined to the
magnetar crust, and have shown that seismological information about
such modes would strongly constrain the physics of the crust 
(Piro 2005, Watts \& Strohmayer 2006, Watts \& Reddy 2007, Samuelsson \& Andersson 2007,
Steiner \& Watts 2009).

However, it was quickly understood that the theoretical analysis of
magnetar oscillations is complicated  by the presence of an ultra
strong magnetic field ($B \sim10^{14} - 10^{15}$ $\rm G$)
that is frozen into the neutron star and penetrates both the
crust and the core. The field provides a channel for an
intense hydro-magnetic interaction
between the motion of the crust and the core, which becomes effective on the timescale of $\ll 1$
second (Levin 2006).
Since the QPOs are observed for hundreds of seconds after the flare,
it is clear that the coupled motion of the crust and the core must be
considered. In recent years, significant theoretical effort has gone into the study of
this problem
(e.g., Glampedakis et al. 2006, Levin 2007, Gruzinov
2008b, Lee 2008). This paper's analysis is based, in part, on an approach of
Levin (2007, L07).

To make progress in computing the coupled crust-core motion,
L07 has studied the time evolution of an axisymmetric toroidal displacement of a
star with axisymmetric poloidal magnetic
field. In that case the Alfven-type motions on different flux surfaces
decouple from each other, a  well-known fact from previous MHD studies
(for a review see Goedbloed \& Poedts 2004, hereafter GP).
One can then formulate the full dynamics of the
system in terms of discrete modes of the crust which are coupled to a
continuum of Alfven modes in the
core. L07 demonstrated that (i) the global
modes with frequencies inside the continuum are strongly damped via a 
phenomenon known in MHD as {\it resonant
absorption} (see GP), and (ii) in many cases, asymptotically the system tends
to oscillate with the frequencies close to the continuum edges. This
result was later confirmed by Gruzinov 2008b, who has
used a powerful analytical technique to solve the L07's normal-mode
problem (Gruzinov noted that
the resonant absorption is mathematically equivalent to Landau damping).
Oscillations near the continuum edge frequencies were also observed in a number of
numerical general-relativistic MHD simulations of purely fluid stars (Sotani et al.~2008,
Colaiuda et al.~2009, Cerda-Duran et al.~2009).

Apart from finding QPOs near the continuum edges, L07's  dynamical simulations 
identified transient 
QPOs with drifting
frequencies; these were transiently amplified near the crustal
frequencies. No explanation for the
origin of the drifts was given.

In this paper, we extend the previous analyses of the hydro-magnetic
crust-core coupling in an essential
way. In section 2,
we re-analyse L07's toy model of a single oscillator coupled to a
continuum, and we show that this
system generically contains the {\it edge normal modes} with frequencies
near the continuum edges.
We show that these modes dominate the late-time dynamics of the system,
and develop a formalism
which allows one to  predict analytically the edge mode's amplitude from
the initial data. We then explore the
effect of viscosity on the system (introduced as a friction between
the neighbouring continuum
oscillators), and show that the edge mode is longer lived than all other
motions of the system. We also
provide a non-trivial analytical formula for the time dependence of the
overall energy dissipation.

In section 3, we describe how transient QPOs, not associated with the 
normal modes of the system, are obtained when parts of the
core spectrum consist of densely and regularly spaced discrete modes
(and in section 5 we show that such an array of discrete modes is expected
when the magnetic field in the core is not perfectly axisymmetric but has some degree
of tangling). As a by-product of our analysis, we
explain the origin of the QPO frequency drifts
seen in L07 simulations. We
provide simple analytical fits to the drifts,
and show that when the
regularity of the continuum sampling is removed (e.g, when the  frequencies are sampled as
random numbers picked from the
continuum range), the drifts disappear.

In section 4, we set up models with a more realistic hydro-magnetic structure of the
neutron-star core.
We show how to find the continuum modes and their coupling to the crust
for an arbitrary axisymmetric
poloidal field, with an arbitrary density profile on the core (L07s
calculations, for simplicity and concreteness,
 were restricted to constant-density
core and homogeneous magnetic field). We treat a more complicated case of
a mixed axisymmetric
toroidal-poloidal field, with radial stratification, in the Appendix B.
We demonstrate
that for realistic field configurations, the Alfven continuum of modes coupled to the crust
may show a
number of gaps. If a crustal
mode frequency belongs to one of these gaps, a strong global discrete mode arises
which dominates the late-time
dynamics and  whose frequency also belongs to
the gap. The frequency of the gap global mode does not generally coincide
with, but is often close to that of the crust.
We suggest that it was these gap modes that appeared in Lee's (2008)
calculations as well-defined discrete
global modes.

So far, only axisymmetric magnetic fields have been considered in the magnetar-QPO
literature, with the Alfven continuum modes occupying the flux surfaces of the field.
In section 5
we argue that if the field is not axisymmetric but instead is highly tangled, then 
the Alfven continuum modes become localized within small regions of individual 
field lines, and therefore become dynamically unimportant. Instead, a set of
discrete Alfven modes appears, with the spacing between the modes strongly dependent
on the degree of field tangling. We devise a phenomenological prescription
which allows us to parametrize the field tangling for computing the dynamically
important modes, and introduce an easily solvable ``square box'' model suitable
for exploring the parameter range.

Finally, in section 6, we use the suite of models built in
the previous sections to explore their connection to the
QPO phenomenology. We find that\newline
(a) within the standard 
magnetar model, it is possible to produce strong long-lived  or transient
QPOs
with frequencies in the range of around $20$---$150$Hz, but only
if the neutrons are decoupled from the Alfven-like motion of the core;
this implies that at least one of the baryonic components of the
core is a quantum fluid.\newline
(b) Our models  could not
produce the
 high-frequency 
$625$Hz QPO within the standard paradigm of a magnetar core
composition.

\section{An oscillator coupled to a continuum: edge modes}

In this section, we study the motion of a harmonic oscillator (which
we hereafter call the large oscillator) which is coupled to a 
continuum of modes.\footnote{In many areas of physics similar models have been 
studied, notably in quantum optics and plasma physics. By contrast with the case studied
here, in these models the range of the continuum frequencies is not limited.}. 
This model was introduced in L07 and it provides a qualitative insight into the
behaviour of crustal modes
(represented by the large oscillator)
coupled to a continuum of Alfven modes in the core of a magnetar. L07 found that if the large
oscillator's proper
frequency was within the range of the continuum frequencies, then the late-time
behaviour of the
system was dominated by oscillatory motion near the edges of the continuum interval.
Here, we give an explanation of this phenomenon in terms of the {\it edge modes}. 
Our analysis allows us to use initial data and 
predict the displacement amplitudes and frequencies of the system at late times.

\begin{figure}[tbp]
\centering
       \includegraphics[width=3in]{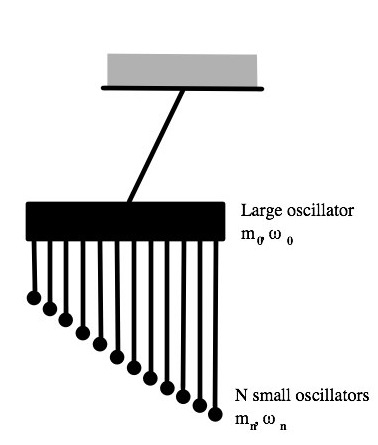}
  \caption{\small Schematic picture of the toy-model. A large number $N$ of small pendulae, 
  representing the (quasi-) continuum, are coupled to one large pendulum, representing the crust.}
    \label{Fig1}
\end{figure}

The model consists of the large mechanical oscillator with mass $M$ and proper
frequency
$\omega_0$, representing a crustal elastic shear mode. Attached to the
large oscillator is a set of $N$ smaller oscillators of mass $m_n$ and proper frequency
$\omega_n$
constituting a 
quasi-continuum of frequencies $\omega_n$ (where $n = 1, 2, ..., N$). The continuum
is achieved
when $N\rightarrow\infty$ while the total small-oscillator mass $\Sigma m_n$ remains
finite.
The convenient pictorial representation is through suspended pendulae, as shown in
Fig. \ref{Fig1} (see also Fig. 2 of L07).  

The equations of motion are obtained as follows.
Each small oscillator is driven by the the motion of the large oscillator:
\begin{eqnarray}
\ddot{x}_n + \omega_n^2 x_n = - \ddot{x}_0
\label{vier}
\end{eqnarray}
where $x_n$ is the displacement of the $n'th$ small oscillator in the frame of
reference 
of the large oscillator, $x_0$ is the displacement of the large oscillator in the
inertial frame of
reference, and the right-hand side represents the non-inertial force acting on
the small oscillator due to the acceleration of the large one.
The large oscillator experiences the combined pull of the small ones:\\
\begin{eqnarray}
M \ddot{x}_0 + M \tilde{\omega}_0^2 x_0 = \sum_i m_i \omega_i^2 x_i
\label{vijf}
\end{eqnarray}
Here  $\tilde{\omega}_0$ is the frequency of the big pendulum corrected for the mass
loading 
by the small pendulae, i.e. $\tilde{\omega}_0^2 = \omega_0^2 \left( M + \sum_i m_i
\right)/M $.

\subsection{Time-dependent behavior.}
In this subsection we explore the behavior of this system by direct numerical
simulations.
We found this to be helpful in the building of our intuition. We defer a semi-analytical
normal-mode
analysis to the next subsection.

We follow L07 and for concreteness concentrate on a specific example; it will be
clear that
the conclusions we reach are general. We choose $\omega_0 = 1$rad/second and mass $M = 1$. 
We choose a total number of 1000 small pendulae with frequencies $\omega_n = (0.5 +
n/1000)$rad/second and 
masses $m_n=m=10^{-4}$, to mimic the continuum frequency range between $0.5$rad/second and
$1.5$rad/second.  
The simulation is initiated by displacing the large oscillator while keeping the
small pendulae relaxed
(this mimics the stresses in the crust), and then releasing. 
The subsequent motion of the system is then followed numerically by using a second order leapfrog 
integration scheme which conserves the energy with high
precision. 
The resulting motion of the large pendulum can be decomposed into three stages (see
Fig. \ref{Fig2A} and Fig. \ref{Fig2B}):

\begin{figure}[tbp]
  \begin{center}
       \includegraphics[width=3.5in]{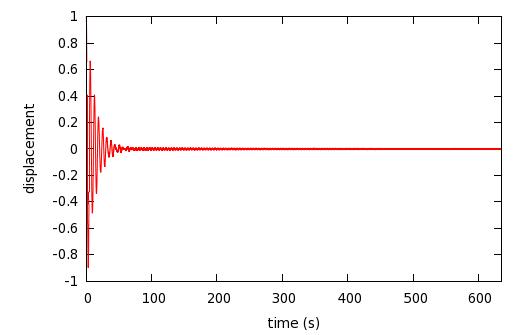}
  \end{center}
  \caption{\small Displacement of the big oscillator as a function of time. }
    \label{Fig2A}
\end{figure}

\begin{figure}[tbp]
  \begin{center}
       \includegraphics[width=3.5in]{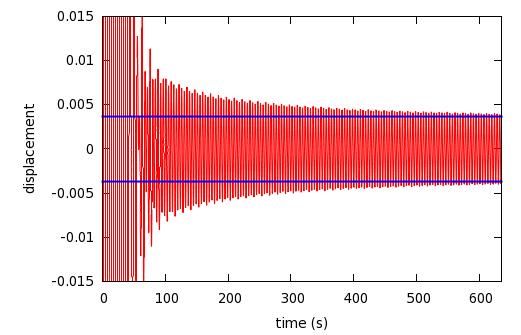}
  \end{center}
  \caption{\small A zoomed-in version of Fig.~\ref{Fig2A}. The blue horizontal lines denote 
  the theoretically predicted amplitude of the dominating upper edge-mode (see section 2.3).}
    \label{Fig2B}
\end{figure}

(1) During the first 50-60 seconds, there is a rapid exponential decay of the large
oscillator's motion, 
during which most of the energy is transferred to the multitude 
(i.e., the 'continuum') of small oscillators.
This is the so-called phenomenon of ``resonant absorption'', which has been studied
for decades in
the MHD and plasma physics community (e.g., Ionson 1978, Hollweg 1987, Goedbloed \& Poedts 2004, L07, 
Gruzinov 2008b). 
In this first stage, the amplitude of the big pendulum motions drops by a factor of $\sim 100$. 

(2) After $\sim 60$ seconds, the exponential decay stops abruptly as the
large oscillator
now reacts to the collective pull of the small ones. This second stage is
characterized by a slow 
algebraic decay of the amplitude of the big pendulum displacement. Gruzinov (2008b)
explains this as being
due to the branch cut in the oscillator's response function.

(3) The motion of the large oscillator stabilizes at a constant level (L07 missed
this stage in his
simulations, which he stopped too early). Fourier transform reveals 2 QPOs
at the frequencies close to the continuum edges, $\omega=0.5$ and $\omega=1.5$; the
same QPO frequencies
can be observed in the previous stage (2) as well.

\begin{figure}[tbp]
  \begin{center}
       \includegraphics[width=3.5in]{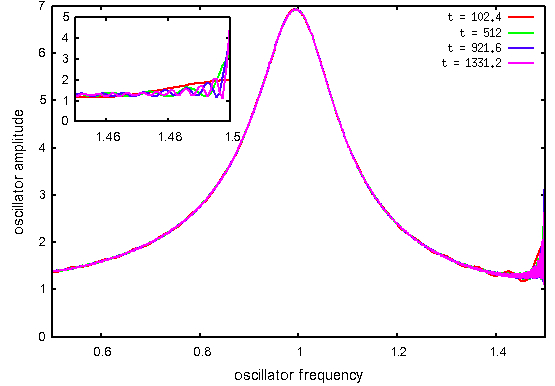}
  \end{center}
  \caption{\small The colored curves show the amplitudes of the small oscillators during the numerical simulation, at different times $t$.}
    \label{Fig5A}
\end{figure}

\begin{figure}[tbp]
  \begin{center}
       \includegraphics[width=3.5in]{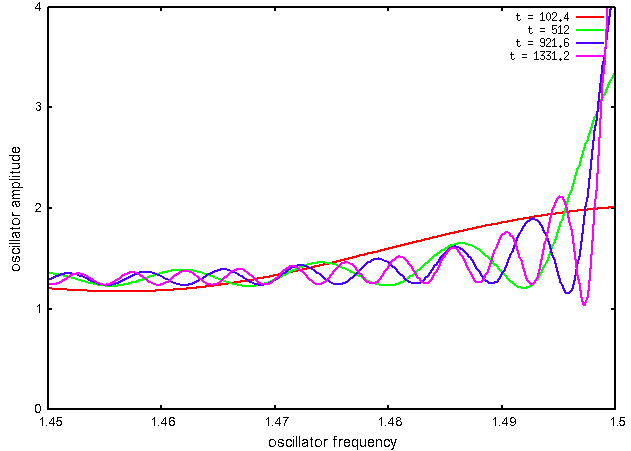}
  \end{center}
  \caption{\small A zoomed-in version of Fig. \ref{Fig5A}. At later times energy is transferred to the oscillators near the edge of the continuum.}
    \label{Fig5AA}
\end{figure}

What is the origin of the QPOs, and how is this eventual stability established? In
Fig.~\ref{Fig5A} and \ref{Fig5AA}, we show how the
amplitude of the small oscillators evolves with time. 

After the initial resonant absorption phase, the amplitude is
distributed as a Lorentzian centered on the frequency around $\omega=1$; this is because the small 
oscillators in resonance with the large one are the ones which gain the most energy.
However, in subsequent
times we see that the energy exchange occurs between the small 
oscillators\footnote{This is much akin to the 
well-known phenomenon of resonant energy exchange between 2 equal-frequency pendulae
hanging on the same
supporting wall.}, and that the net result of this
exchange is the energy flow towards the oscillators whose frequencies are near the
edges. By the time the
third stage begins, the amplitudes of the oscillators near the edge 
stabilize and their phases become locked.
They are pulling and pushing the large oscillator in unison. In the next subsection, 
we show that this behavior
is due to the presence of the {\it edge normal modes}, 
and we shall derive their frequencies and amplitudes.

\subsection{Finding eigenmodes}
In this section we deal with the system of coupled harmonic oscillators, and one
should be able to find
its normal modes using the standard techniques (Landau and Lifshitz mechanics, \S 23). 
However, the fact that all small oscillators are attached to the large one, and
there is 
no direct coupling between
the small oscillators, allows us a significant shortcut (in Appendix A, we treat a
more general problem of {\it several} large oscillators coupled to a multitude of
the core modes).  
We proceed as follows:

Suppose that we impose on the large oscillator a periodic motion with  angular
frequency $\Omega$, 
by driving
it externally with the force $F_{\rm ext}=F_0(\Omega) \exp(i\Omega t)$. This motion
in turn drives 
the small oscillators
according to Eq.~(\ref{vier}):
\begin{equation}
\ddot{x}_n + \omega_n^2 x_n = \Omega^2 x_0,
\label{no6}
\end{equation}
which has the steady state solution:
\begin{equation}
x_n = \frac{\Omega^2}{\omega_n^2 - \Omega^2} x_0 
\label{no7}
\end{equation}
where we have omitted the time dependent factor $\exp(i \Omega t)$ on both sides.
The combined force 
$f_{\rm cont}$ of the small oscillators acting back on the large one (see
Eq.~(\ref{vijf})) is given by
\begin{equation}
f_{\rm cont} \left( \Omega \right) = \sum_n m_n \omega_n^2
\frac{\Omega^2}{\omega_n^2 - \Omega^2} x_0.
\label{no8}
\end{equation}
According to Newton's second law,
\begin{equation}
F_0\left(\Omega\right)+f_{\rm cont}\left(\Omega\right)=-M(\Omega^2-\omega_0^2) x_0.
\label{secondlaw}
\end{equation}
If $\Omega$ corresponds to the normal-mode frequency, then $F_0(\Omega)=0$. Hence by
substituting
Eq.~(\ref{no8}) into Eq.~(\ref{secondlaw}) we get the following eigenvalue equation
for $\Omega$:

\begin{equation}
G(\Omega)=M \left( \omega_0^2 - \Omega^2\right)-\sum_n 
m_n \omega_n^2 \frac{\Omega^2}{\omega_n^2 - \Omega^2}=0.
\label{discretemodes}
\end{equation}

In the continuum limit $N\rightarrow\infty$, the above equation becomes
\begin{equation}
G(\Omega)=M \left( \omega_0^2 - \Omega^2\right)-\int_{\omega_{\rm min}}^{\omega_{\rm
max}} d\omega 
\rho(\omega) \omega^2 \frac{\Omega^2}{\omega^2 - \Omega^2}=0,
\label{contmodes}
\end{equation}
where $\rho(\omega)=dm/d\omega$ is the mass per unit frequency of the continuum modes. 

In the discrete case, the solutions of Eq.~(\ref{discretemodes}) 
are $N-1$ frequencies $\Omega_i$ that are within the quasi-continuum 
($\omega_i < \Omega_{i+1} < \omega_{i+1}$, for $i=1,2,...N-1$; 'quasi-continuum
modes'), and 
$2$ modes with frequencies $\Omega_{\rm low}$ and $\Omega_{\rm high}$ (the
'edge-modes') 
that are near the edges, but outside, of the continuum 
(i.e. $\Omega_{\rm low} < \omega_1$ and $\Omega_{\rm high}>\omega_N$).
It can be shown from Eq.~(\ref{discretemodes})  
that in the limit $N\gg 1$ and $m_n\ll M$, 
the contribution of the small oscillator to the $i$-th quasi-continuum mode is
completely dominated 
by the pendulae that are in close resonance with the mode. More precisely, one can
show that as the 
number of oscillators $N$ increases and $m_n$ decreases, the number of small
oscillators 
contributing to the mode energy remains constant. However,
for the two edge modes there is no such singular behavior in the limit of large
$N$, and 
consequently they play a special role in the dynamics of the system.

This last point is clearly
 seen in the continuum case represented by Eq.~(\ref{contmodes}). The eigenvalue
equation
has no real solutions in the range of small-oscillator continuum $\omega_{\rm
min}<\Omega<\omega_{\rm max}$,
since the response function $G(\Omega)$ is ill-defined in this
interval\footnote{There is a complex
solution if the integration in the expression for $G(\Omega)$ is performed along the
contour chosen
according to the Landau rule. One then obtains a ``resonantly absorbed'' or
``Landau-damped'' mode 
(Gruzinov 2008b, L07), which exactly represents the exponential decay of stage (1) in
our numerical 
experiment of the previous subsection.}. However, the edge modes on both sides of
the continuum
interval remain, and their frequencies can be found by numerically evaluating the
zero points 
of $G(\Omega)$ in Eq.~(\ref{contmodes}). For the numerical of the previous
subsection, one finds
$\Omega_{\rm low}= 0.5 - 8.2 \cdotp 10^{-6}$
and $\Omega_{\rm high}=1.5 + 8.6 \cdotp 10^{-4}$. 
Analytically, one can find the following scaling for the distance $\delta \omega_{\rm
edge}$ between the mode frequency and the nearest edge $\omega_{\rm edge}$ of the
continuum range:

\begin{equation}
 {\delta \omega_{\rm edge}\over \omega_{\rm
edge}}=C\exp\left\{-{M|\Omega_0^2-\omega_{\rm edge}^2|\over 
\rho(\omega_{\rm edge})\omega_{\rm edge}^3}\right\},
\label{distance}
\end{equation}
where $C$ is a constant of order unity. The larger is the density of continuum modes
at the 
edge $\rho(\omega_{\rm edge})$, the further is the edge mode pushed away from the
continuum  range.
It is particularly interesting to consider the case when the continuum interval is
limited
by a turning point (L07) with the divergent density of states near the edge,
$\rho(\omega)=A/\sqrt{|\omega-\omega_{\rm edge}|}$, where $A$ is a constant. 
In this case the distance from the edge-mode
frequency to the nearest continuum edge is given by
\begin{equation}
{\delta \omega_{\rm edge}\over \omega_{\rm edge}}=C\left\{ 
{A\omega_{\rm edge}^{7/2}\over M|\Omega_0^2-\omega_{\rm edge}^2|}\right\}^2.
\label{distance1}
\end{equation}
The quadratic dependence in Eq.~(\ref{distance1}) vs. the exponential dependence in
Eq.~(\ref{distance}) implies that the continua with turning points typically feature 
much more pronounced edge modes and stronger QPOs than the ones with linear edges.
In the next section, we show how to calculate the edge-mode amplitudes and
QPO strengths from the initial data.


%

\subsection{Late time behavior of the system}

In the continuum limit, the only modes with real oscillatory frequency are the edge
modes. Thus,
as we demonstrate explicitly below,  they  dominate the late-time dynamics of the
system when the
number $N$ of small oscillators becomes large. Our analysis proceeds as
follows:\newline
Lets define a new set of variables, expressed as a vector $\vec{X}$ with components
 $X_{0} = \sqrt{M}x_{0}$ and $X_{n} = \sqrt{m_n} \left( x_{0} + x_{n} \right)$ for
$n=1,..., N$.
With these new variables, the kinetic energy of the system is a trivial quadratic
expression
\begin{equation}
K={1\over 2}\dot{\vec{X}}\cdot\dot{\vec{X}},
\label{K}
\end{equation}
where the inner product of two vectors $\vec{A}$ and $\vec{B}$ is 
defined as $\vec{A}\cdot\vec{B}=\Sigma_{j=0}^NA_jB_j$. The potential
energy is a positive-definite quadratic form, whose exact form is unimportant here.
The mutually orthogonal eigenmodes $\vec{X}^i$ 
can be found via a procedure outlined in the previous
section\footnote{Alternatively, they can be
found by diagonalizing the potential-energy quadratic form.}. 
Their eigenfreguencies $\Omega_i$ are identified by finding zeros of $G(\Omega)$ in
Eq.~(\ref{discretemodes}), and the corresponding eigenvector components are given by 
\begin{eqnarray}
X^i_0&=&1\label{eigen1}\\
X^i_n&=&{\omega_n^2\over \Omega_i^2-\omega_n^2}\nonumber
\end{eqnarray}, 

Lets assume that we initiate our simulation by displacing the large oscillator by an
amount $x_0\left( 0 \right)$ 
while keeping the small oscillators relaxed $x_n \left( 0 \right) = 0$ and 
all initial velocities at zero. 
In the new variables, the initial state of the system is given by the vector
$\vec{X}(0)$, 
where $\vec{X}_0=\sqrt{M}x_0(0)$ and
$\vec{X}_n=\sqrt{m_n}x_0(0)$. The time evolution of the system is given by:
\begin{equation}
\vec{X}(t)=
\Sigma_{\Omega_i}\cos(\Omega_i
t)\left(\vec{X}^i\cdot\vec{X}^i\right)^{-1}\left(\vec{X}(0)\cdot\vec{X}^i\right)
\vec{X}^i. 
\label{no10}
\end{equation}
Substituting the initial conditions, and the expression in Eq.~(\ref{eigen1}) for
the eigenvector components,
we get
\begin{equation}
\vec{X}(t)=
\sum_{\Omega_i}{\cos(\Omega_i t){M+\sum_{n}{m_n\omega_n^2\over
\omega_n^2-\Omega_i^2}\over
M+\sum_{n}{m_n\omega_n^4\over (\omega_n^2-\Omega_i^2)^2}}} \vec{X}^i. 
\label{no10}
\end{equation}
The coordinate of the large oscillator is simply given by
$x_0(t)=\vec{X}_0(t)/\sqrt{M}$.

For the continuum of small modes, the above expansion breaks down, since the 
eigenvalue equation
has no real solutions inside the continuum range. 
However, the edge modes are well defined, and they determine the dynamics at late
times.
Therefore, for the continuum case we can still write down the analogous expression
which is valid only at late times:
\begin{eqnarray}
\vec{X}(t) = \Sigma_{\Omega_{\rm edge}}\cos(\Omega_{\rm edge} t)
\frac{\vec{X}(0)\cdot\vec{X}_{\rm edge}}{\vec{X}_{\rm edge}\cdot\vec{X}_{\rm edge}}
\vec{X}_{\rm edge}
\label{edgedec}
\end{eqnarray}
The sums of  Eq.~(\ref{no10}) are replaced with the corresponding integrals,
and we have the following expression for the displacement of the large oscillator at
late times:
\begin{equation}
x_0(t)=x_0(0)\sum_{\Omega_{\rm edge}} \cos(\Omega_{\rm edge}t) {M+\int{d\omega \rho(\omega){\omega_n^2\over
\omega_n^2-\Omega_{\rm edge}^2}}
\over M+\int{d\omega \rho(\omega){\omega_n^4\over (\omega_n^2-\Omega_{\rm
edge}^2)^2}}}
\label{latex0}
\end{equation}
This expression is in excellent agreement with the numerical simulations. In the
numerical example 
of subsection $2.1$, the upper edge mode dominates the late-time behavior of the
system, and
its calculated contribution is plotted in Fig.~\ref{Fig2B}, together with the numerically
simulated motion.

\subsection{The effect of viscosity}
We now add  an extra degree of realism by introducing viscous friction into the system.
In MHD, continuum modes
are spatially localized, and the effect of viscosity is to frictionally couple the
neighboring modes (see,
e.g., Hollweg 1987). In our simple model we introduce viscosity by adding frictional
forces between the 
neighboring oscillators,
\begin{equation}
f_{n,n+1}=-f_{n+1,n}=\gamma (\dot{x}_n-\dot{x}_{n+1}),
\label{friction1}
\end{equation}
where $f_{n,n+1}$ is the force from the $n$'th oscillator acting on the $(n+1)$'th. 

\begin{figure}[tbp]
  \begin{center}
       \includegraphics[width=3.5in]{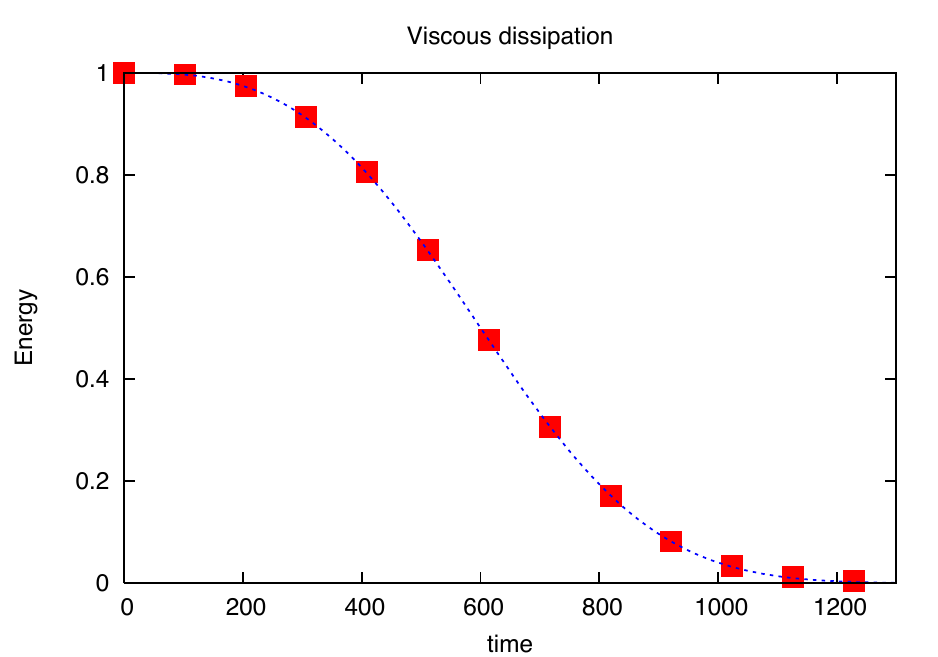}
  \end{center}
  \caption{\small The red squares show the viscous dissipation of the total energy during the numerical simulation. 
  The dotted blue curve shows the analytical solution from Eq.~(\ref{Ediss})  }
    \label{Fig4}
\end{figure}

We now calculate how the system dissipates energy as a function of time. We will
show that it 
occurs in two stages (see Fig. \ref{Fig4}): (1) Initially, the small oscillators are
strongly and 
simultaneously excited by the "Landau-damped" large oscillator,
then they become dephased, with the average relative motion between the neighboring 
oscillators growing linearly in time. This leads to a very 
rapid dissipation of the bulk of the initial energy. (2) The edge modes persist,
since the participating small oscillators move in phase and the energy dissipation
is small. 
The energy of the modes is damped exponentially on
a timescale much longer than that of the first stage.

The dissipated energy is given by
\begin{equation}
W_{\rm diss}=\Sigma_{n=1}^{N-1} \gamma (\dot{x}_{n+1}-\dot{x}_{n})^2.
\label{wdiss}
\end{equation}

In the continuum limit, the small oscillators are labeled not by a discrete index
$n$, but
by a continuous variable $\lambda$. The expression for the dissipated energy is
then
\begin{equation}
W_{\rm diss}=\int d\lambda\tilde{\gamma} 
\left({\partial^2 x_{\lambda}(t)/\over \partial\lambda \partial t}\right)^2,
\label{wdisscont}
\end{equation}
where $\tilde{\gamma}$ is the viscous coefficient. After the initial exponential damping
of the
large oscillator and the excitation of the small
oscillators, the latter initially move independently, with
\begin{equation}
x_\lambda(t)\simeq\tilde{x}(\lambda)\cos[\omega_\lambda t],
\label{xlambda}
\end{equation}
where $\tilde{x}(\lambda)$ is the amplitude of the $\lambda$'th oscillator.
From the above equation, we then obtain
\begin{equation}
\left\langle\left({\partial^2 x_{\lambda}(t)/\over \partial\lambda \partial
t}\right)^2\right\rangle=
{1\over 2} \left\{[d(\tilde{x}_\lambda\omega_\lambda)/d\lambda]^2+\omega_\lambda^2
\tilde{x}_\lambda^2(d\omega_\lambda/d\lambda)^2t^2\right],
\label{dlambda}
\end{equation}
where the $\langle ...\rangle$ stands for time-averaging over many oscillation periods.
For times $t\gg d\log x_\lambda/d\omega_\lambda$ the second term on the right-hand side
of Eq.~(\ref{dlambda}) dominates.  For a simple model with
$d\omega_\lambda/d\lambda=const$
and $\rho(\omega)=const$,
\begin{equation}
dE/dt\propto -At^2E,
\label{dEdt}
\end{equation}
where $E$ is the total energy of the system and
$A=(\tilde{\gamma}/\rho)(d\omega_\lambda/d\lambda)$. 
The analytical solution for the energy and dissipated power,
\begin{eqnarray}
E&=&E_0\exp\left(-{1\over 3}At^3\right),\label{Ediss}\\
W_{\rm diss}&=&-{dE\over dt}=At^2E_0\exp\left(-{1\over 3}At^3\right)\nonumber
\end{eqnarray}
agrees very well with numerical simulations, see Fig. \ref{Fig4}. While the equations above
were derived
for restrictive assumptions  ($d\omega_\lambda/d\lambda=const$
and $\rho(\omega)=const$), we found that the analytical formulae in
Eq.~(\ref{Ediss}) provide a
good fit for a large variety of simulations. This is because it is the small
oscillators with the 
frequencies 
near that of the large oscillator which carry most of the energy, and in that narrow
band our   
approximations hold.

\begin{figure}[tbp]
  \begin{center}
       \includegraphics[width=3.5in]{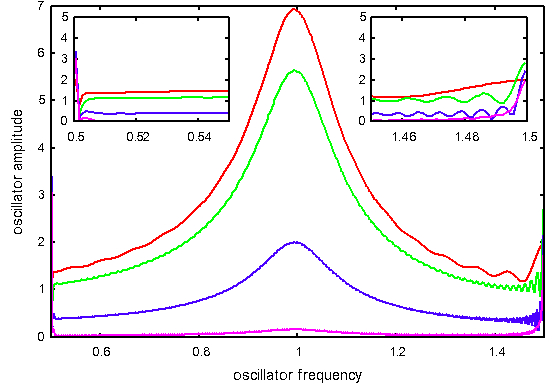}
  \end{center}
  \caption{\small As in Fig.~(\ref{Fig5A}), this figure shows the amplitudes of the small oscillators at different times $t$. 
  The energy of most oscillators is drained due to viscous dissipation. At late times, only the oscillators near the 
  edges of the continuum have substantial energy.}
    \label{Fig5B}
\end{figure}

After the energy dissipation due to dephasing is over, only the edge modes remain.
This is illustrated in Fig. \ref{Fig5B}, where we show how the energies of the small
oscillators evolve
with time. At late times, only the oscillators taking part in the edge modes move
substantially;
this is because they remain in phase and do not dissipate much. At this stage the
energy is drained
by slow exponential decay of the edge modes. 

\section{Transient and drifting QPOs}
Finite-size MHD systems feature a mix of continuum and discrete modes (see Poedts et al., 1985 and GP). 
For axisymmetric field configurations the continuum modes occupy the whole 
 flux
surfaces and play an important role in the oscillatory dynamics; this was the motivation for
L07 and our study of the previous section. We will argue in 
section 5 that if the core field is highly tangled, the continuum modes become 
localized in space and discrete core modes will play a more
important role. Thus it is important to study the case when the the crustal modes
are coupled to a set of discrete core
modes.  In this section we  show that if the frequencies of the discrete modes are
regularly spaced in some frequency intervals, then the system displays
{\it transient QPOs} that  are entirely missed by its normal-mode analysis. This is
interesting from the observational point
of view, since many of the observed magnetar QPO features are transient. 

\begin{figure}[tbp]
  \begin{center}
      \includegraphics[width=3.5in]{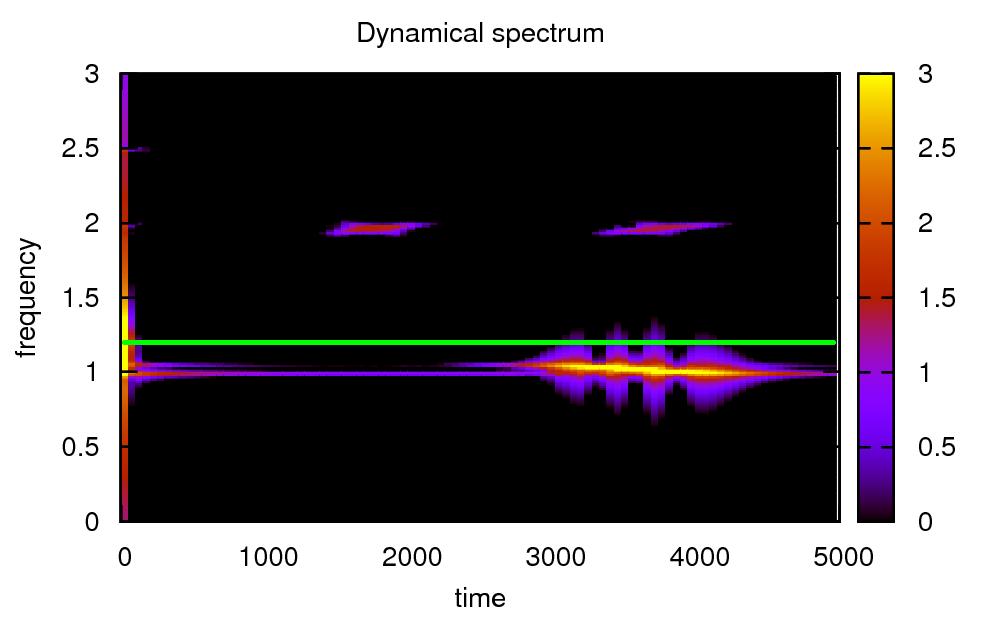}
  \end{center}
  \caption{\small Dynamical spectrum from a simulation where we have designed the  continuum so as to produce transient QPO's at frequencies $\omega = 1$ and $\omega = 2$ (the colored scale denotes $\log (\rm{power})$). The green horizontal line denotes the frequency of the large oscillator ($\Omega = 1.2$). }
    \label{Figx}
\end{figure}

\begin{figure}[tbp]
  \begin{center}
       \includegraphics[width=3.5in]{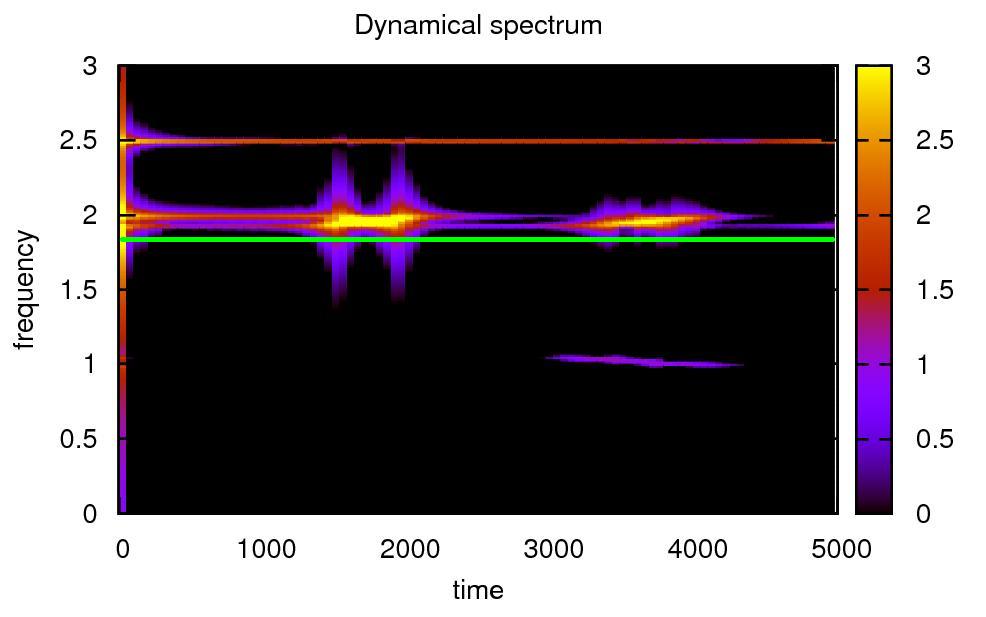}
  \end{center}
  \caption{\small Dynamical spectrum from a simulation with a continuum that is identical to the one of Fig \ref{Figx}. We have shifted the frequency of the large oscillator, denoted by the green horizontal line, to $\Omega = 1.8$. By comparison wit Fig. \ref{Figx} it is clear that the drifting QPO's at $\omega = 2$ are now much stronger as they are closer to the large oscillator frequency. 
Note that the edge mode at $\omega = 2.5$ is clearly visible.}
    \label{Figy} 
\end{figure}

Suppose that a set of discrete modes are located in the interval $\Delta \omega$
around frequency $\omega_0$ and are separated by a regular frequency 
interval $\delta\omega$, and assume the following  hierarchy:
\begin{equation}
\delta\omega\ll\Delta\omega\ll\omega_0.
\label{hierarchy}
\end{equation}
After the modes are excited,  they are initially in phase but will de-phase rapidly
on the timescale $1/\Delta\omega$. However, at times $t_n=2\pi n/\delta\omega$
the modes come into phase again and pull coherently on  the large oscillator.
Therefore, a transient QPO feature 
should appear around these times at a frequency close
to $\omega_0$. In Fig. \ref{Figx} and Fig. \ref{Figy} we show the dynamical spectrum from a simulation where the
conditions for transient QPOs were engineered for two 
frequencies. The transient QPOs agree well with the expectations. As is seen from the figures, the
strongest transient QPOs are those whose frequencies
are the closest to that of the large oscillator; this is because the response of the
large oscillator is the strongest around its proper frequency.

\begin{figure}[tbp]
  \begin{center}
       \includegraphics[width=3.5in]{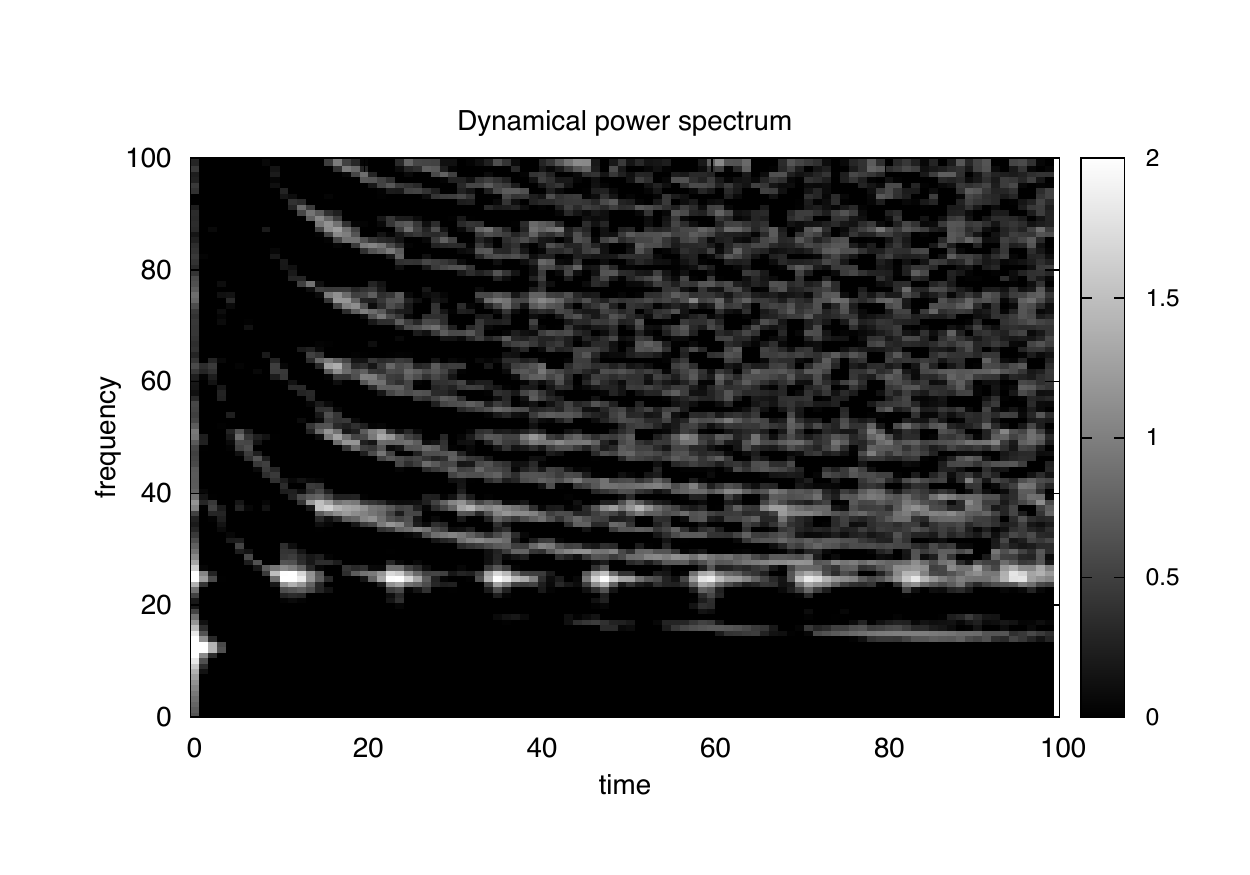}
  \end{center}
  \caption{\small Dynamical power spectrum of the spherical magnetar model from L07. The gray scale denotes $\log(\rm{power})$.}
    \label{Figw}
\end{figure}

\begin{figure}[tbp]
  \begin{center}
      \includegraphics[width=3.5in]{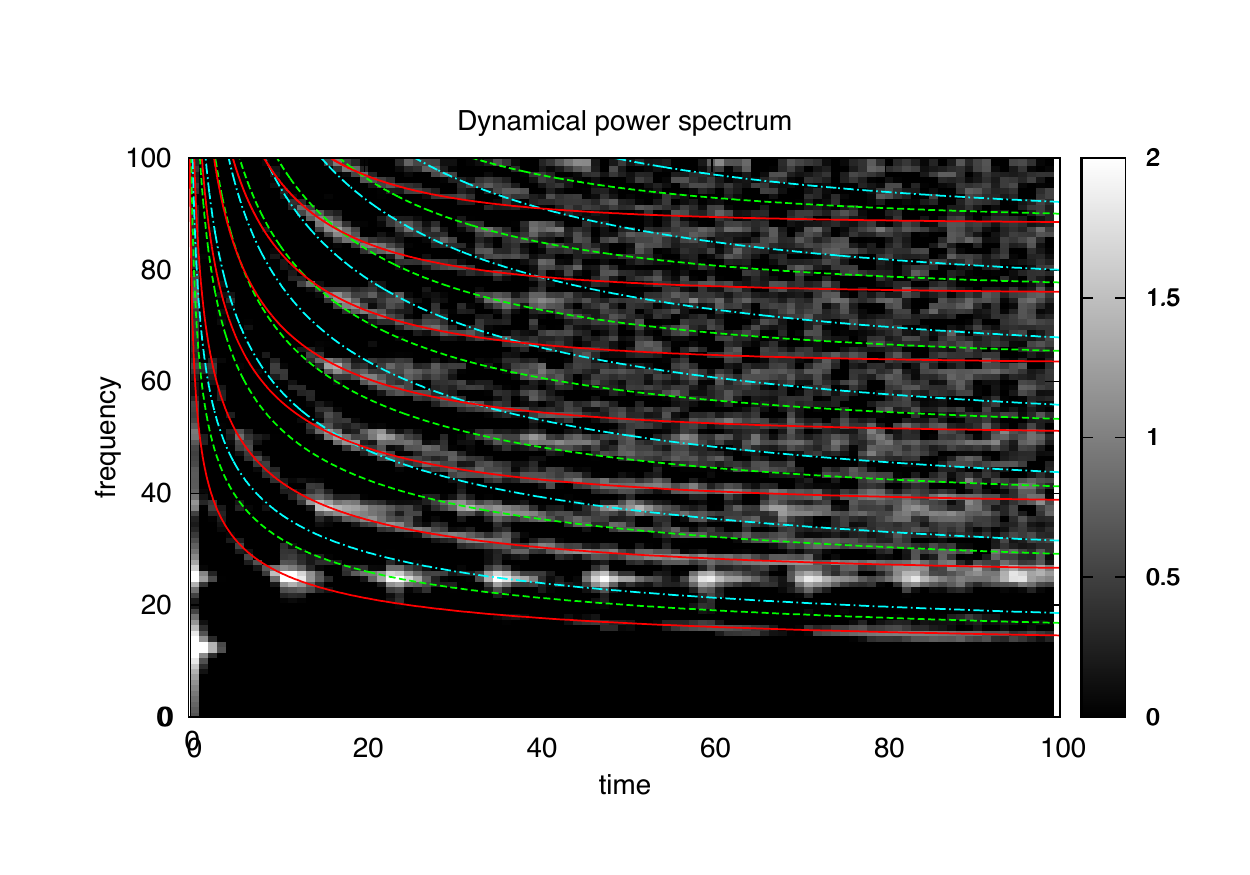}
  \end{center}
  \caption{\small We have used Eq.~(\ref{delta_t}) to fit the drifting QPO's from figure \ref{Figw}. 
  The red curves are  $n=1$ drifts, green curves are $n=2$ and blue curves are $n=3$. The higher frequency drifts originate from Alfven overtones.}
    \label{Figz}
\end{figure}

One can now easily understand  the frequency drifts in Fig. 10 of L07 (Fig. \ref{Figw} in this paper). In the
simulations of that paper, the core continuum was sampled with a set of densely and 
regularly-placed Alfven 
modes by   slicing the field into finite-width flux shells. The spacing
$\delta\omega$ between the modes was not constant but a function of the Alfven
frequency
$\omega$. In that case, the QPO drifts with the QPO frequency $\omega(t)$ given by
the inverse relation
\begin{equation}
t(\omega)={2\pi n\over\delta\omega(\omega)}.
\label{delta_t}
\end{equation}
With this relation we are able to fit all of L07 drifting QPOs, as shown in Fig. \ref{Figw} and \ref{Figz}.
Note that multiple QPOs correspond to different branches of
the Alfven continuum. As expected, the drifting QPOs amplified near the crustal
frequencies, since there the response of the crust to the
core modes' pull is the strongest. 

\section{More realistic magnetar models}

In this section we extend the constant magnetic field and constant-density
magnetar model from L07 to include more realistic pressure and density profiles 
and more general (but still axisymmetric) magnetic field configurations. Our aim is
to use this model to: 1) calculate numerically Alfven eigenmodes and their eigenfrequencies on
different flux surfaces inside the star, in order to determine the continuous
spectrum of the fluid core, and 2) use these modes to simulate the dynamics of a
realistic magnetar. 
In order to calculate the Alfven eigenmodes and eigenfrequencies for a realistic
magnetar model, we employ the linearized equations of motion for an axisymmetric
magnetized, self-gravitating plasma. The general equations, which are derived in
detail in Poedts et al.~(1985, hereafter P85) and given in their equations (53) and
(54), constitute a fourth order system of coupled ordinary differential equations in
the case of a mixed poloidal and toroidal magnetic field. The formalism of P85 is briefly summarized
in the Appendix B. In the case of a purely
poloidal magnetic field, the system simplifies to two uncoupled second order
differential equations (P85, equations (70) and (71)).  \\

\subsection{The model}

We assume our star is non-rotating and neglect  its deformation due to the
magnetic pressure, which is expected to be small. 
Therefore, we consider a spherically symmetric background model that is a solution
of the Tolman-Oppenheimer-Volkoff equation (TOV equation). The hydrostatic
equilibrium is calculated using a SLy equation of state (Douchin \& Haensel 2001;
Haensel \& Potekhin, 2004; Haensel, Potekhin \& Yakovlev 2007), which can be found
in tabulated form on the website \textit{http://www.ioffe.ru/astro/NSG/NSEOS/}. The
integration of the TOV equation is performed using a 4th order Runge-Kutta scheme,
integrating from the center of the star outward until we reach a mass density $\rho
= 1.3 \cdotp 10^{14}$ $\rm g/cm^3$, which is consistent with the crust-core
interface in the equation of state from Douchin \& Haensel (2001). The resulting
model has a central mass density $\rho_c = 10^{15}$ $\rm g/cm^3$, a total mass of
$1.40$ $\rm M_{\odot}$ and a radius of $R_{\rm core} = 1.07 \cdotp 10^6$ $\rm cm$. 
To this spherical model we add a poloidal magnetic field, which we generate by
placing a circular current loop of radius $a$ and current $I$ around the center of
the star. The field is singular near the current loop, however
all the field lines which connect to the crust (and thus are physically related to observable 
oscillations) carry finite field values.
This particular field configuration is chosen as an example; there is an infinite
number of ways to generate poloidal field configurations. In the appendix B we will
add to this field a toroidal component and calculate the corresponding Alfven
continuum of the core. 

\subsection{The continuum}
In order to find the equations of motion for the magnetized material in the neutron
star core, we would  need to add self-gravity 
to the ideal magnetohydrodynamic equations. This problem has been solved by P85 in a
tour the force mathematical approach. 
In that paper the authors assume a self-gravitating axisymmetric equilibrium with
a field geometry consisting of mixed poloidal and toroidal field 
components, and they derive linearized equations of motion. For this field geometry
it is convenient to work with so-called flux-coordinates $(\psi, \chi, \phi)$. 
The basic concept behind this curvilinear coordinate system is the magnetic
flux-surface, which is defined as the surface perpendicular to the 
Lorentz force $\vec{F}_L \propto \vec{j} \times \vec{B}$. From this definition it is
clear that the magnetic field lines lie in flux surfaces. If
one considers a closed loop on a flux surface which makes one revolution around the
axis of symmetry, then the magnetic  flux $\psi$ through the
loop depends on the flux surface only and is the same for all of the loops.
Therefore $\psi$ is chosen as the coordinate labeling the flux surfaces. 
In each flux-surface we can denote a position by its azimuthal angle $\phi$ and its
'poloidal' coordinate $\chi$, which is defined as the length along $\phi=const$
line. 
In P85, it is shown that the equations of motion allow for a class of oscillatory solutions that
are located on singular flux surfaces, constituting a continuum of eigenmodes and
eigenfrequencies. In the case of a purely poloidal field ($B = B_{\chi}$), the continuum
solutions are degenerate and polarized either parallel ($\xi_{\chi}$) or perpendicular
($\xi_{\phi}$) to the magnetic field lines. 
In the latter case the displacement is $\phi$-independent.
It is clear that in contrast to the $\chi$-polarized modes, the $\phi$-polarized
modes are purely horizontal and are therefore unaffected by gravity. 
This latter case is considered here.

The equation of motion is in this case simply the Alfven wave equation:
\begin{equation}
\frac{\partial^2 \xi_{\phi}(\psi,\chi)}{\partial t^2} = F\left[\xi_\phi(\psi,\chi)\right],
\label{poedts}
\end{equation}
where the operator $F$ is given by
\begin{equation}
F\left[\xi_\phi(\psi,\chi)\right]=
\frac{B}{4 \pi x \rho}
\frac{\partial}{\partial \chi} \left[ x^2 B \frac{\partial}{\partial \chi} \left(
\frac{\xi_{\phi}(\psi,
\chi)}{x} \right) \right].
\label{poedts1}
\end{equation}
Here $x$ is the distance to the magnetic axis of symmetry. Although in the presence
of a mixed poloidal and toroidal field the equations still give rise to a continuous
set of solutions, 
the calculations are significantly complicated as the continuum modes are affected
by the  toroidal component of the field, by gravity, and by compressibility. For the
sake of simplicity we will ignore toroidal fields in our dynamic simulations. We
will however, calculate the continuum frequencies for a mixed poloidal and toroidal
field in the Appendix B.

\begin{figure}[tbp]
  \begin{center}
       \includegraphics[width=3.5in]{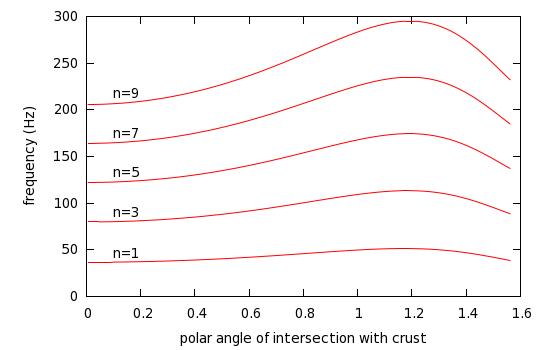}
  \end{center}
  \caption{\small The red curves show the Alfven frequencies $\sigma_n$ as a function of the angle $\theta (\psi)$, the polar angle at which the flux-surface $\psi$ intersects the crust. Since we are only considering odd crustal modes, the only Alfven modes that couple to the motion of the star are the ones with an odd harmonic number $n$. This particular continuum was calculated using a poloidal field with an average surface value $B_{\rm surface} \sim 6 \cdotp 10^{14}~ \rm{G}$, generated by a circular ring current of radius $a = R_{*}/2$.}
    \label{core_cont1}
\end{figure}

For determining the spectrum of the core continuum, the appropriate
boundary conditions are $\xi_{\phi}(\chi=\chi_c)=0$, where $\chi_c(\phi)$ marks the
location of the crust-core interface.
With this boundary condition, Equation (\ref{poedts}) constitutes a Sturm-Liouville
problem on each separate flux surface $\psi$. 
Using the stellar structure model and magnetic field configuration from section 4.1, 
we can calculate the eigenfunctions and eigenfrequencies for each flux
surface $\psi$. The reflection symmetry of the stellar model and the magnetic field
with respect to the equatorial plane assures  that the eigenfunctions of
Eq.~(\ref{poedts}) are either symmetric or anti-symmetric with respect to the
equatorial plane. We can therefore determine the eigenfunctions by integrating
Eq.~(\ref{poedts}) along the magnetic field lines from the equatorial plane $\chi =
0$ to the crust-core interface $\chi = \chi_c \left( \psi \right)$. Let us consider
the odd modes here for which $\xi_{\phi} \left( 0 \right) = 0$, and solve
Eq.~(\ref{poedts}) with the boundary condition $\xi_{\phi} \left(\chi_c \right) =
0$ at the crust-core interface; for even modes, the boundary condition is
$d\xi_{\phi} \left( 0 \right)/d\chi=0$. We find the eigenfunctions by means of a
shooting method; using fourth order Runge-Kutta integration we integrate from $\chi
= 0$ to $\chi = \chi_c$. The correct eigenvalues $\sigma_n$ and eigenfunctions
$\xi_n \left( \chi \right)$ are found by changing the value of $\sigma$ until the
boundary condition at $\xi_n$ is satisfied. In this way we gradually increase the
value of $\sigma$ until the desired number of harmonics is obtained. In figure \ref{core_cont1}
we show a typical resulting core-continuum.

\begin{figure}[tbp]
  \begin{center}
       \includegraphics[width=3.5in]{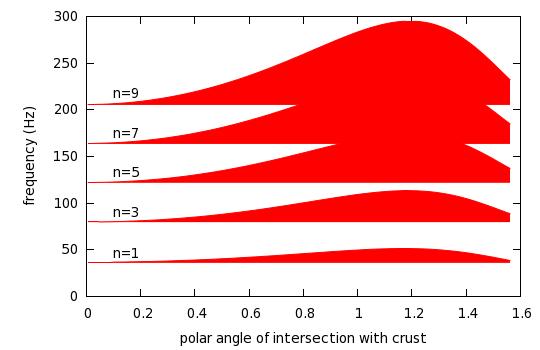}
  \end{center}
  \caption{\small After filling the curves from Fig. \ref{core_cont1}, 'gaps' in the continuum become visible around $\sigma \sim 70 ~\rm Hz$ and $\sigma \sim 120 ~\rm Hz$.}
    \label{core_cont2}
\end{figure}
 
According to Sturm-Liouville theory the normalized eigenfunctions $\xi_n$ of
Eq.~(\ref{poedts}) form an orthonormal basis with respect to the following inner
product:
\begin{eqnarray}
\langle \xi_m, \xi_n \rangle=
\int_0^{\chi_c} r\left( \chi \right) \xi_m\left( \chi \right) \xi_n\left( \chi
\right) d \chi = \delta_{m,n}
\label{eq_inner}
\end{eqnarray}
Where $\delta_{m,n}$ is the Kronecker delta and $r = 4 \pi \rho / B_{\chi}$ is the
weight function. We have checked that 
the solutions we find 
satisfy the orthogonality relations.

We are now ready to compute the coupled crust-core motion. Here we follow L07 and
assume that the crust is an infinitely thin elastic 
shell\footnote{It is straightforward to relax this assumption, and carry out the
analysis of this section for the finite  crustal thickness.  However, 
from Section 2 it is clear that the interesting dynamics is dominated by the
spectral structure of the core Alfven waves; therefore
in order to flesh out the physics we choose the simplified model of the crust.}. We
label the 
lattitudinal location   by the flux surface $\psi$ intersecting the crust, and consider
the crustal axisymmetric displacements $\bar{\xi}_\phi(\psi)$. 
In the MHD approximation, the  magnetic stresses enforce a no-slip boundary
condition at the crust-core interface, such that $\xi_{\phi} \left( \psi, \chi_c
\right) = \bar{\xi}_{\phi} \left( \psi, \chi_c \right)$ instead of $\xi_{\phi}
\left( \psi, \chi_c \right) = 0$. It is useful to make the following substitution
\begin{eqnarray}
\zeta \left( \psi,\chi \right) \equiv \xi_{\phi} \left( \psi,\chi \right) -
\bar{\xi}_{\phi} \left( \psi \right) w\left( \psi,\chi \right)
\label{substit0}
\end{eqnarray}
where we choose the function $w\left( \psi,\chi \right)$ so that (a) it corresponds
to the static displacement in the core and hence satisfies
 $F \left(w\left( \psi,\chi \right) \right) = 0$, and (2)  $w\left( \psi,\chi_c
\right)=1$. 
Therefore the new quantity satisfies the boundary condition $\zeta \left(
\psi,\chi_c \right)=0$ and can be expanded into
the Alfven normal modes $\xi_n$ which satisfy the same boundary conditions.

We now proceed by substituting Eq.~(\ref{substit0}) into Eq.~(\ref{poedts}) 
thus obtaining a simple equation of motion for $\zeta$
\begin{eqnarray}
\frac{\partial^2 \zeta \left( \psi,\chi \right)}{\partial t^2} - F \left( \zeta
\left( \psi,\chi \right) \right) = - w\left( \psi,\chi \right) \frac{\partial^2
\bar{\xi}_{\phi} \left( \psi \right)}{\partial t^2}
\label{substit1}
\end{eqnarray}
From the definition of the operator $F$ it follows that for the odd modes
\begin{eqnarray}
w \left( \psi, \chi \right) = x \left( \psi, \chi \right) \int_0^{\chi} \frac{K
\left( \psi \right)}{x^2 \left( \psi, \chi' \right) B_{\chi} \left( \psi, \chi'
\right)} d \chi'.
\label{substit1a}
\end{eqnarray}
Here the constant $K \left( \psi \right)$ is chosen such that $w\left( \psi,\chi_c
\right) = 1$, in order that $\zeta = 0$ on both boundaries. We  expand $\zeta$ and
$w$ into a series of $\xi_n$'s: 
\begin{eqnarray}
\zeta \left( \psi,\chi,t \right) = \sum_n a_n \left( \psi,t \right) \xi_n \left(
\psi,\chi \right)\\
w \left( \psi,\chi \right) = \sum_n b_n \left( \psi \right) \xi_n \left( \psi,\chi
\right)
\label{substit2}
\end{eqnarray}
Eq.~(\ref{substit1}) reduces  to the following equations of motion for the eigenmode
amplitudes $a_n$
\begin{eqnarray}
\frac{\partial^2 a_n \left( \psi \right)}{\partial t^2} + \sigma_n^2 \left( \psi
\right) a_n \left( \psi \right) = - b_n \left( \psi \right) \frac{\partial^2
\bar{\xi}_{\phi}}{\partial t^2}
\label{substit3}
\end{eqnarray}
These equations show how the core Alfven modes are driven by the motion of the
crust. To close the system,
we must address the motion of the crust driven by the hydromagnetic pull from the core.

The equation of motion for the crust is  given by
\begin{eqnarray}
\frac{\partial^2 \bar{\xi}_{\phi}}{\partial t^2} = L_{\rm el} \left(
\bar{\xi}_{\phi} \right) + L_B
\label{eqa}
\end{eqnarray}
Where the acceleration due to elastic stresses $L_{\rm el}$ is
\begin{eqnarray}
L_{\rm el} \left( \bar{\xi}_{\phi} \right) = \omega_{\rm el}^2 \left[
\frac{\partial^2 \bar{\xi}_{\phi}}{\partial \theta^2} + \cot{\left(\theta \right)}
\frac{\partial \bar{\xi}_{\phi}}{\partial \theta} - \left( \cot{\left(\theta
\right)}^2 - 1 \right) \bar{\xi}_{\phi} \right]
\label{eqb}
\end{eqnarray}
where $\theta$ is the polar angle (cf.~L07). The acceleration $L_B$ due to the
magnetic stresses between the crust and the core can be expressed as
\begin{eqnarray}
L_B = - \frac{x B^2 }{4 \pi \Sigma} \cos{\alpha } \frac{\partial}{\partial \chi}
\left( \frac{\xi_{\phi} }{x} \right)_{\chi = \chi_{\rm crust}}
\label{substit4}
\end{eqnarray}
where $x$ is the distance to the axis of the star, $\Sigma$ is column mass-density
of the crust and $\alpha$ is the angle between the magnetic field line and the
normal vector of the crust.

It is convenient to express the crustal displacement $\bar{\xi}_{\phi}$ as a Fourier
series, being a sum normal modes of the free-crust problem. Using Eq.~(\ref{eqb}) is
straightforward to show analytically that the eigenfunctions $f_l$ of the free-crust
problem (Eq.~(\ref{eqa}) with $L_B = 0$) are
\begin{eqnarray}
f_l \left( \theta \right) \propto \frac{dY_{l0} \left( \theta \right)}{d\theta}
\label{eq0}
\end{eqnarray}
with eigenfrequencies
\begin{eqnarray}
\omega_l =  \omega_{\rm el} \sqrt{\left( l-1 \right) \left( l+2 \right)}
\label{eq1}
\end{eqnarray} 
Here $Y_{l0}$ is the $m=0$ spherical harmonic of degree $l$. The normalized
functions $f_l$ form an orthonormal basis, so that
\begin{eqnarray}
\int_0^{\infty} f_l \left( \theta \right) f_m \left( \theta \right) \sin{\left(
\theta \right)} d\theta = \delta_{l,m}
\label{eqc}
\end{eqnarray}
where $\delta_{l,m}$ is again the Kronecker delta. The crustal displacement can then
be expressed in terms of $f_l$
\begin{eqnarray}
\bar{\xi}_{\phi} \left( \theta, t \right) = \sum_l c_l \left( t \right) f_l \left(
\theta \right)
\label{eqc}
\end{eqnarray}
Substituting Eq.~(\ref{eqc}) into Eq.~(\ref{eqa}) we obtain the equations of motion
for the crustal mode amplitudes $c_l$
\begin{eqnarray}
\frac{\partial^2 c_l}{\partial t^2} + \omega_l^2 c_l = \int_0^{\pi} L_B \left(
\theta, t \right) f_l \left( \theta \right) \sin \theta d\theta
\label{substit5}
\end{eqnarray}
We can express $L_B$ as

\begin{eqnarray}
L_B \left( \psi, t \right) =
-\frac{B_{\chi}^2 \left( \psi \right)}{4 \pi \Sigma}
\cos{\left( \alpha \left( \psi \right) \right)} \left[ \sum_n a_n \left( t \right) 
\frac{\partial \xi_n \left( \psi \right)}{\partial \chi}  \right. \\ \nonumber
\left. + \frac{K \left( \psi \right)}{x \left( \psi \right) B \left( \psi \right)} \sum_k c_k \left( t \right)
f_k \left( \theta \left( \psi \right) \right)\right]_{\chi = \chi_c} 
\label{substit6}
\end{eqnarray}

Up to this point the derived equations of motion for the crust and the fluid core
are exact. We are now ready to discretize the continuum by converting the integral
of Eq.~(\ref{substit5}) into a sum over $N$ points $\theta_i$. In order to avoid the
effect of phase coherence (see section 3) which caused   drifts in the results
from L07, we sample the continuum randomly over the $\theta$-interval $\left[ 0,
\pi/2 \right]$. In the following, functional dependence of the coordinate $\psi$ or
$\theta \left( \psi \right)$ is substituted by the discrete index $i$ which denotes
the $i$-th flux surface. 

\begin{eqnarray}
\frac{\partial^2 c_l}{\partial t^2} + \omega_l^2 c_l = 2\sum_i L_B \left( \theta_i,
t \right) f_{il} \sin \theta_i \Delta \theta_i \\ \nonumber
= - \sum_i \sin{\left( \theta_i \right)} \Delta \theta_i f_{il} \left[
\frac{B_{\chi, i}^2}{2 \pi \Sigma} \cos{\left( \alpha_i \right)} \left( \sum_n
a_{in} \frac{\partial \xi_{in}}{\partial \chi}  \right. \right. \\ \nonumber
\left. \left. + \frac{K_i}{x_i B_{\chi, i}} \sum_k c_k f_{ik} \right) \right]_{\chi = \chi_c}
\label{substit7}
\end{eqnarray}

\begin{eqnarray}
\frac{\partial^2 a_{in}}{\partial t^2} + \sigma_{in}^2 a_{in} = - b_{in} \sum_l
\frac{\partial^2 c_l}{\partial t^2} f_{il}
\label{subs3}
\end{eqnarray}

These are the equations that fully describe dynamics of our magnetar model. As with
the toy model from section 2 we integrate them using a second order leap-frog
scheme which conserves the total energy to high precision. As a test we keep track
of the total energy of the system during the simulations. Further we have checked
our results by integrating equations (\ref{substit7}) and (\ref{subs3}) with the
fourth-order Runge-Kutta scheme and found good agreement with leap-frog integration.

\subsection{Results}
Based on our section-2 results, we have a good idea of what type of dynamical behavior
 should occur in our  more realistic magnetar model. 
First, we expect that crustal modes with frequencies inside the Alfven continuum
will be damped quickly by resonant absorption ("Landau-damping" in the terminology
of Gruzinov 2008b).
 Second, as with our previous model we expect the late time behavior of the system
to show QPO's near the edges of the continuum, or edge modes. The third 
 important feature of our model is that the continuum may possibly contain gaps,  as
is shown in 
Fig. \ref{core_cont2}. In this case there is the possibility that crustal frequencies fall
inside the gaps and remain undamped. In all of our simulations these expectations
have come true. We will now show the results from a simulation which illustrate the
above mentioned effects. 

The basic freedom of choice that we have for our model is the strength and geometry
of the equilibrium magnetic field. We choose here a purely poloidal magnetic field
with an average strength at the surface of $B_{\rm surf} = 10^{15}~\rm G$, induced
by a circular current loop of radius $a=0.5R_*$. This field gives us a gap in the
continuum at frequencies $53 < \omega < 78  ~\rm Hz$. 

We consider the lowest degree odd crustal modes with frequencies $\omega_2 = 40~\rm
Hz$ and $\omega_4 = 84.5~\rm Hz$, which we couple to 5000 continuum oscillators (the
Alfven continuum). We sample the continuum at 1000 randomly chosen flux surfaces,
and at each flux surface we consider 5 Alfven overtones.

Like with our toy model from section 2, we initiate the simulation by displacing
the crust ($c_2 = c_4 = 1$) while keeping the continuum oscillators (the Alfven
modes) relaxed ($a_{in}=0$). 

\begin{figure}[tbp]
  \begin{center}
    \includegraphics[width=3.5in]{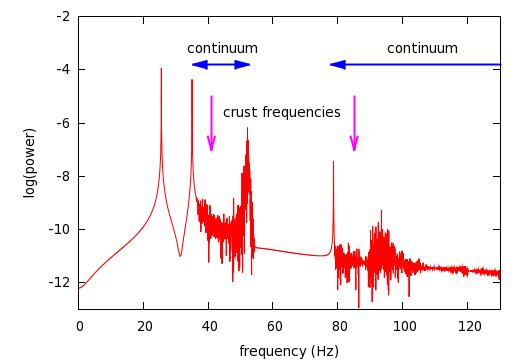}
  \end{center}
  \caption{\small Power spectrum of the crustal dynamics for a magnetar with a single 'gap' in the 
Alfven continuum. In this case the crustal frequencies are within the continuum, 
causing the crust modes to be Landau-damped.}
    \label{Fig9}
\end{figure}

\begin{figure}[tbp]
  \begin{center}
    \includegraphics[width=3.5in]{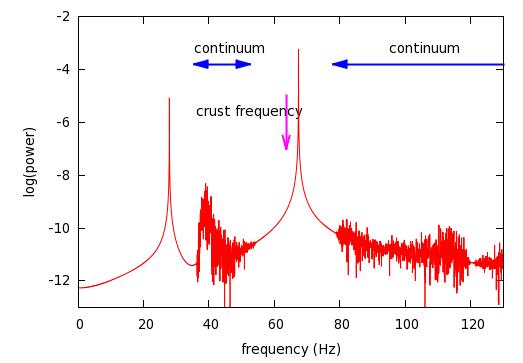}
  \end{center}
  \caption{\small Power spectrum of the crustal dynamics for a magnetar with a single 'gap' in the Alfven continuum. 
  The global mode within the gap is not damped, and 
  its frequency is similar, but not identical, to that of the crustal mode in the same gap.}
    \label{Fig10}
\end{figure}

In Figures \ref{Fig9} and \ref{Fig10} we show the resulting power spectrum for 2 different models. In the first one,
the crustal frequencies are located inside the core continuum range, and the peaks due to
the edge modes appear. By contrast, in the second case one of the crustal frequencies belongs to
the gap, and a peak representing the global gap mode strongly stands our above the background.
We note that the gap-mode's frequency lies close to but does not coincide with the crustal-mode frequency;
we found this to be a generic feature of our models, with the difference of  $~10$\% for the typical
model parameters. 
The gap modes are particularly interesting because they have relatively large amplitudes, 
and are not as strongly damped
by viscosity as the edge modes.

It must be emphasized that for all persistent modes in the system, the position in the frequency space
of the core Alfven continuum plays the key role in setting the global-mode frequency and in determining 
its longevity.

We note that Lee (2008) has used a different method to identify discrete modes in a
magnetar with
similar magnetic configuration to ours. These modes were not associated with crustal
frequencies, and
we strongly suspect that they were located in the  gaps of the continuum spectrum
and could be identified with the
edge or gap modes presented in this work.

\section{Tangled magnetic fields}
Our preceding discussion of the continuum was predicated on the foliation of the
axisymmetric magnetic
field into the flux surfaces, with each of the singular continuum mode localized on
the flux surfaces.
These modes are "large"-they are coherent over the spacial extent comparable to the
size of the system,
and thus they play an important role in the overall dynamics-they are responsible for
the resonant absorption of the crust oscillations, and contribute to generating the
edge and gap modes.
But what happens if the field cannot be foliated into the flux surfaces, but is
instead tangled in
a complicated way? One can argue that the continuum part of the spectrum
still persists, as follows:

\begin{figure}[tbp]
  \begin{center}
     \includegraphics[width=3in]{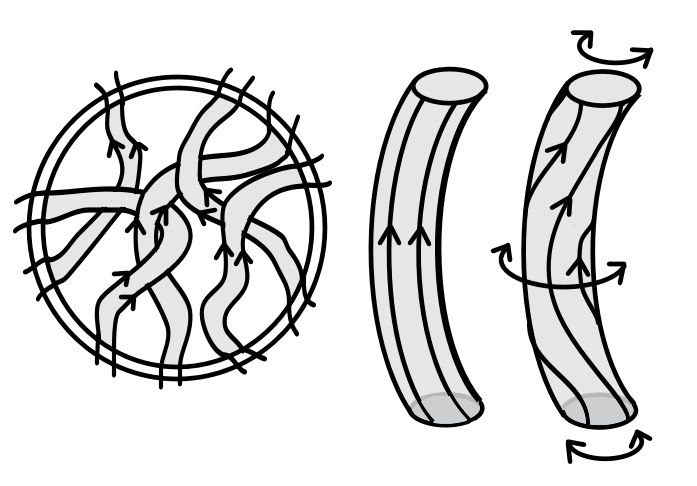}
  \end{center}
  \caption{\small Schematic illustration of tangled a magnetic field inside a magnetar.
   Locally, the field consists of flux tubes which contain a continuum of twisting Alfven modes.}
    \label{Fig7}
\end{figure}

Consider an arbitrary field line anchored at the crust-core interface at both ends,
and choose a tube of
field lines of infinitesimal radius which is centered on the original field line
(see Fig. \ref{Fig7}). It is clear
that a twisting Alfven mode exists in the tube: it is 
obtained by the circular rotation of the fluid around the central field line,
propagating along the central field line with the local Alfven velocity. Since there
is a continuum of the
field-line lengths, there is also a continuum of Alfven modes. However, the modes we
constructed are highly
localized in space and and have a small leverage in the overall dynamics. We
conjecture that the more
tangled the fields are, the less role do the singular  continuum modes play in the
overall dynamics.

Whilst we cannot rigorously prove this conjecture, we can motivate it as follows:
consider an area element $\delta S$ of random orientation with the normal $\hat{n}$
 inside the star, and consider a shearing motion along the element. This shearing
motion will be resisted
by the $B_{\hat{n}}$ component of the magnetic field, with the effective shear
modulus of order
\begin{equation}
\mu_{\rm eff}\sim {\langle B_{\hat{n}}^2\rangle \over 4\pi},
\label{mueff}
\end{equation}
where $\langle ... \rangle$ stands for averaging over the area element. 
For ordered field, it is possible to choose the orientation of the area element so
that 
$\mu_{\rm eff}\simeq 0$; the presence of such an orientation makes a fundamental
difference between 
MHD and elasticity theory and is responsible for the presence of continuous spectrum
in MHD. However,
if the linear size of the $\delta S$ is greater than the characteristic length on
which the field is
tangled, then $\mu_{\rm eff}$ is non-zero for all orientations of $\hat{n}$.
Therefore, for highly-tangled
fields there can be no large-scale singular continuum modes, and their existence is
restricted to the small scales.
Hence our assertion that for strongly tangled fields continuum modes play a
secondary dynamical role. 

One is then faced with the problem when crustal modes are coupled to a set of discrete
core Alfven modes. In Appendix A we show how to find the eigen-solution
of  such a problem, provided that
all of the coupling coefficients are known.

How does one quantify the degree to which the fields are tangled? 
Some insight comes from the numerical simulations of Braithwaite and colleagues,
who have studied what type of fossil fields remain in a stratified star after an
initial period of fast relaxation.
Consider a stable
fossil field
field configuration, such as
the one obtained in the simulations of Braithwaite and Spruit (2004) and Braithwaite and Nordlund (2006) [see also
Gruzinov (2008a) for analytical  considerations].
There,
the final field is nearly, but not perfectly axisymmetric and has a small-scale random
component. For a less-centrally concentrated initial field,
Braithwaite (2008) shows that the final fossil field is in general non-axisymmetric and can have
a complicated topology.\footnote{Gruzinov (2009) demonstrates that even this situation is not the most general. He finds that the
relaxed field generally has multiple current sheets, and argues that the global field relaxation is dominated by the
dissipation  within these singular layers. The details do not concern us for the purposes of this paper.} 

As a starting point, we shall consider the nearly axisymmetric field with a small random
component. The latter acts like a small extra shear
modulus $\mu_{\rm eff}$ and 
dynamically couples the flux surfaces of the axisymmetric component.  We then quantify the
degree of tangling by the relative value
of $\mu_{\rm eff}$ and $B^2/(4\pi)$.

\subsection{simple model: "square" neutron star}
\begin{figure}[tbp]
  \begin{center}
     \includegraphics[width=3in]{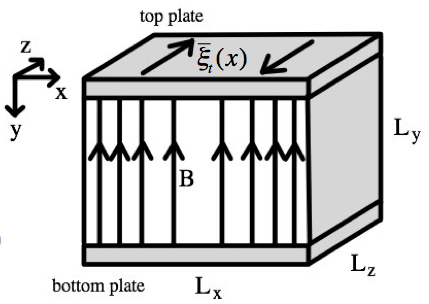}
  \end{center}
  \caption{\small Schematic illustration of the box model. Perfectly conducting incompressible
fluid is sandwiched between perfectly conducting top and bottom
plates. The box is periodic in $z$-direction
and the displacements of the plates (crust) are in the $z$-direction. The magnetic field is directed along $y$-axis and its strength varies as a function of $x$.}
    \label{boxje}
\end{figure}

To study this idea further, we specify a very simple model of a neutron star, motivated by the one
considered in Levin (2006, hereafter
L06) see Fig. \ref{boxje} that  never-the-less captures the essential physics.

Consider a perfectly conducting homogeneous fluid of density $\rho$
contained in a box with width $L_x$, length
$L_y$ and depth $L_z$. The magnetic field in this box is everywhere aligned with the
$y$-axis and its strength is a function of $x$ only. We assume that gravity is zero
and consider a Lagrangian displacement $\xi \left( x,y,t \right)$ of the fluid along
the $z$-direction; we specify
periodic boundary conditions in this direction (one should think of the $z$ direction as azimuthal). 
We now add to this model a small effective shear modulus $\mu_{\rm eff}$ due to the field tangling.
The fluid equation of motion is:
\begin{equation}
\frac{\partial^2 \xi}{\partial t^2} = c_A^2\left(x\right)\frac{\partial^2
\xi}{\partial y^2} + c_s^2\nabla^2 \xi
\label{waveeq}
\end{equation}
Here $c_A \left( x \right)$ is the Alfven velocity and $c_s=\sqrt{\mu_{\rm eff} / \rho}$ is the $\mu_{\rm eff}$-generated
 shear velocity. If we assume a small shear speed, i.e. $c_s << c_A$,
Eq.~(\ref{waveeq}) reduces to\\
\begin{equation}
\frac{\partial^2 \xi}{\partial t^2} = c_A^2\left(x\right)\frac{\partial^2
\xi}{\partial y^2} +c_s^2 \frac{\partial^2 \xi}{\partial x^2}.
\label{waveeqred}
\end{equation}
We now find the core Alfven eigenmodes. After adapting the no-slip boundary conditions\\
\begin{eqnarray}
\xi \left( -\frac{L_x}{2},y,t \right) = \xi \left( \frac{L_x}{2},y,t\right) = 0,
\nonumber\\
\xi \left( x,-\frac{L_y}{2},t \right) = \xi \left( x, \frac{L_y}{2},t\right) = 0,
\label{boundarycons}
\end{eqnarray}
the problem can be easily solved by separation of variables $\xi \left( x, y,
t\right) \propto e^{i\omega t} \sin \left\{\pi m [(y / L_y)+1/2] \right\} X \left( x
\right)$, where $m=1,2,...$. Plugging this in Eq.~(\ref{waveeqred}) we find for the
the $x-$dependent part of the solution: \\
\begin{eqnarray}
c_s^2\frac{\partial^2 X}{\partial x^2} = \left[ \omega^2 - \omega_{A,m}^2 \left( x
\right)\right]X.
\label{solution}
\end{eqnarray}
Here $\omega_{A,m}\left(x\right) = \pi m c_A\left(x\right)/L_y$ can be interpreted
as the frequency of the $m$-th Alfven overtone at $x$. From the above expression it
is clear that in the limit of very small $c_s$, the solution for $X$ must be close
to zero everywhere except in a very small neighborhood of $\omega_{A,m} (x) =
\omega$. It is in this limit that the solutions are located on singular flux
surfaces. However, in the presence of the non-vanishing shear velocity $c_s$, the
eigenmodes spread out on neighboring field lines, effectively coupling the motion on
different flux surfaces. The continuum of Alfven frequencies $\omega_{A,m} \left( x
\right)$ will in this case be no longer solutions of the system. Instead, the
coupling term gives rise to a discrete set of solutions rather than a continuum.
Eq.~(\ref{solution}) is the mathematical equivalent of 
Schr\"{o}dingers equation, which can in general cases be solved numerically. 
However, for many special case exact solutions exist. Let us consider, for the sake
of simplicity, a field configuration in our box such that:
\begin{eqnarray}
c_A^2 \left( x \right) = a_{c_A} x^2 + c_{A,0}^2
\label{c_A}
\end{eqnarray}
We can rewrite Eq.~(\ref{solution}) as follows: 
\begin{eqnarray}
c_s^2 \frac{\partial^2 X}{\partial x^2} = -\frac{\pi^2 m^2 a_{c_A}}{L_y}x^2 X +
\left(\omega_m^2 - \frac{\pi^2 m^2 c_{A,0}^2}{L_y}\right)X
\label{schrod}
\end{eqnarray}
This differential equation is the mathematical equivalent of the quantum harmonic
oscillator problem for which the exact solution is well known. The eigenfrequencies
are given by 
\begin{eqnarray}
\omega_{mn}^2 = \pi \left( 1+2n \right) m c_s \sqrt{a_{c_A}}/L_y + c_{A,0}^2 \pi^2
m^2/L_y.
\label{sol}
\end{eqnarray}
Here $n$ $(=0,1,...)$ is the 'quantum' number labeling the harmonic-oscillator wavefunctions. We see that instead of a continuum, we obtain a densely packed
discrete set of frequencies with the frequency spacing $\omega_{m,n} - \omega_{m,n-1}
\sim \pi m c_s \sqrt{a_{c_A}}/L_y\omega_{m,n}$. 

\begin{figure}[tbp]
  \begin{center}
     \includegraphics[width=3.5in]{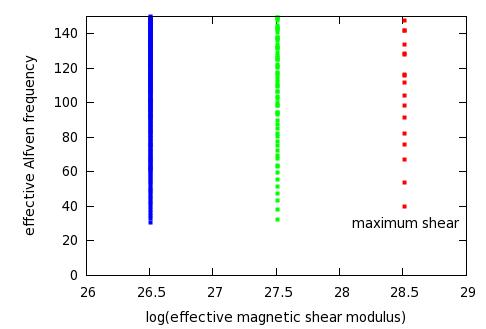}
  \end{center}
  \caption{\small Alfven frequencies as a function of the effective magnetic shear modulus. As one decreases the shear, the spectrum tends to a continuum.}
    \label{Fig11}
\end{figure}

With the no-slip boundary conditions on the left and right sides $x=\pm L_x/2$, the eigenvalue equation must be solved numerically. An example of such calculation is
shown in Fig. \ref{Fig11}. There, the spacing between the discrete Alfven modes is shown to increase as one increases the effect of the field tangling
characterized by the $\mu_{\rm eff}$.

We now introduce the crustal modes into the problem by making the top and bottom of the box elastic and mobile.
We allow their displacement $\bar{\xi}_{t,b}(x,t)$ in the z-direction, and impose the boundary conditions on the sides:
\begin{equation}
\bar{\xi}_{t,b}(-L_x/2,t)= \bar{\xi}_{t,b}(L_x/2,t)=0.
\label{boundary}
\end{equation}
Here the subscripts "$t$" and "$b$" stand for the top and bottom of the box, respectively. The top and bottom are assumed to be thin
and have mass $M_{\rm cr}$ and surface density $\sigma=M_{\rm cr}/(L_xL_z)$ . The crustal equation of motion is given by
\begin{eqnarray}
{\partial^2 \bar{\xi}_{t}\over \partial t^2}&=&v_s^2{\partial ^2 \bar{\xi}_{t}\over \partial x^2}-{\{B_zB_x\}_t\over 4\pi\sigma}\nonumber\\
{\partial^2 \bar{\xi}_{b}\over \partial t^2}&=&v_s^2{\partial ^2 \bar{\xi}_{b}\over \partial x^2}+{\{B_zB_x\}_b\over 4\pi\sigma},
\label{crustmotion}
\end{eqnarray}
where $v_s$ is the shear velocity in the crust. The crustal angular frequencies are given by $\omega^{\rm cr}_j=j\pi v_s/L_y$ with the corresponding
crustal mode eigenfunctions $\bar{\xi}_j=\sin\{j\pi [(x/L_x)+1/2]\}$ , where $j=1,2,...$ is roughly equivalent to $l$ in the spherical case.
The symmetry of the problem allows one to consider either symmetric 
$\bar{\xi}_{t}=\bar{\xi}_b$ or antisymmetric $\bar{\xi}_{t}=-\bar{\xi}_b$  crustal modes. This will couple to the symmetric ($m=1, 3, 5, ...$) or
antisymmetric ($m=2, 4, 6, ...$) Alfven modes of the core.

Just as in section 4, it is now convenient to define a new variable $\zeta(x,y,t)$ 
for the core displacement:
\begin{equation}
\zeta(x,y,t) = \xi(x,y,t) - \xi_0(x,y,t),
\label{tildexi}
\end{equation}
where
\begin{eqnarray}
\xi_0(x,y,t)&=&{1\over 2}\left(\bar{\xi}_t(x,t)+\bar{\xi}_b(x,t)\right)\nonumber\\
&+&\left(\bar{\xi}_t(x,t)-\bar{\xi}_b(x,t)\right){y\over L_y}.
\label{newxi}
\end{eqnarray}
The new variable observes the regular boundary condition $\zeta=0$ on all the box edge, and  satisfies the following inhomogeneous partial differential 
equation:
\begin{equation}
\left({\partial^2\over \partial t^2}-c_A^2(x){\partial^2\over \partial y^2}-c_s^2{\partial^2\over \partial x^2}\right)\zeta(x,y,t)=g(x,y,t),
\label{eqtilde}
\end{equation}
where 
\begin{equation}
g(x,y,t)=-\left({\partial^2 \over \partial t^2}-c_s^2{\partial^2\over \partial x^2}\right)\xi_0(x,y,t).
\label{driver}
\end{equation}
The advantage of the new variable is that it satisfies the regular boundary condition $\zeta=0$ on all the boundaries of the box.
It can therefore be expanded as a series consisting of eigenfunctions $\xi_{mn}$ of the right-hand side of Equation (\ref{waveeqred}):
\begin{equation}
\zeta(x,y,t)=\Sigma_{mn} a_{mn}(t) \xi_{mn}(x,y).
\label{expansioncore}
\end{equation}
The rest of the procedure is very similar to that in section 4. We expand the crustal displacement into a series consisting of the eigenmode
wavefunctions $\bar{\xi}_j$:
\begin{eqnarray}
\bar{\xi}_t(x,t)&=&\Sigma_j p_j(t) \bar{\xi}_j(x)\nonumber\\
\bar{\xi}_b(x,t)&=&\Sigma_j q_j(t) \bar{\xi}_j(x),\label{crustaldosp}
\end{eqnarray}
where $p_j(t)$ and $q_j(t)$ are real numbers. The magnetar deformation is now fully represented by a set of
generalized coordinates $[p_j(t), q_j(t), a_{mn}(t)]$. The coupled equations of motion are derived by following the procedure specified
 in section 4. We obtain the following system of equations:
\begin{eqnarray}
\ddot{a}_{mn}+\omega_{mn}^2 a_{mn}&=&-\Sigma_j\left[\ddot{p}_j+c_s^2\left({j\pi\over L_x}\right)^2p_j\right]
                                                                                        \alpha^{(p)}_{(mn)j} \nonumber\\
                                                                         &  &-\Sigma_j\left[\ddot{q}_j+c_s^2\left({j\pi\over L_x}\right)^2q_j\right]
                                                                                        \alpha^{(q)}_{(mn)j},\label{couple1}
\end{eqnarray}
and
\begin{eqnarray}
\ddot{p}_j+{\omega^{\rm cr}_j}^2p_j&=&-{\rho c_A^2\over \sigma}
\Sigma_{mn} \beta_{j(mn)} a_{mn}\label{couple2}\\
\ddot{q}_j+{\omega^{\rm cr}_j}^2q_j&=&-{\rho c_A^2\over \sigma}
\Sigma_{mn}(-1)^{m+1}\beta_{j(mn)} a_{mn}\nonumber
\end{eqnarray}
where
\begin{eqnarray}
\alpha^{(p)}_{(mn)j}&=&{\int \left({1\over 2}+{y\over L_y}\right)
 \xi_{mn}(x,y)\bar{\xi}_j(x) dx dy\over \int [\xi_{mn}(x,y)]^2 dx dy}\nonumber\\
\alpha^{(q)}_{(mn)j}&=&{\int \left({1\over 2}-{y\over L_y}\right)
 \xi_{mn}(x,y)\bar{\xi}_j(x) dx dy\over \int [\xi_{mn}(x,y)]^2 dx dy}\label{couple4}
\end{eqnarray}
and
\begin{equation}
\beta_{j(mn)}={\int \left({\partial \xi_{mn}(x,y)\over \partial y}\right)_{y=L_y/2}
                        \bar{\xi}_j(x) dx\over \int [\bar{\xi}_j(x)]^2 dx}.
\label{couple5}
\end{equation}
Thus we have obtained a system of linear second-order differential equations, which describes the time evolution of the square-box magnetar. These equations are solved by trancating
all the series [i.e., rescricting the range of indices $(m, n, j)$] and then by either solving the eigenvalue
problem in order to find the normal modes, or by integrating the equations numerically\footnote{Our favored method here
is again the energy-conserving second-order leapfrog. It is both fast and stable over long integration times.}.
One then checks that the series trancation does not introduce errors in the magnetar's motion within the frequency range of our interest.

So far we have worked in the approximation of the thin crust, i.e. we have effectively included the crustal modes
which have no radial nodes in their wavefunction. However, several observed high-frequency QPOs,
and in particular the strong QPO at $625$Hz (Watts \& Strohmayer 2006) necessitate introduction
of higher radial-order modes into our model. In the square-box model we do this phenomenologically,
as follows. We assume that higher radial-order crustal modes have  amplitudes $p_{sj}(t)$ and
$q_{sj}(t)$ and natural eigenfrequencies $\omega^{\rm cr}_{sj}$, with $s$ being the number of radial nodes, and assume that they cause displacement at the
crust-core interface given by $\bar{\xi}_j(x)$. This mirrors realistic spherically-symmetric case where
the functional form of the crust-core displacement due to the torsional $\nabla \times Y_{lm}$ mode of the
$n$'th radial order is
a very weak function $n$.  The amplitudes $p_{sj}(t)$ and
$q_{sj}(t)$ are then introduced on into the equations of motions (\ref{couple1}) and (\ref{couple2})
in the same way as the other $p_j$ and $q_j$ amplitudes, with the same $j$-dependent coupling
coefficients but with $\omega^{\rm cr} _{sj}$ instead of $\omega^{\rm cr}_j$ on the left-hand side of Eq.~(\ref{couple2}).

\begin{figure}[tbp]
  \begin{center}
         \includegraphics[width=3.5in]{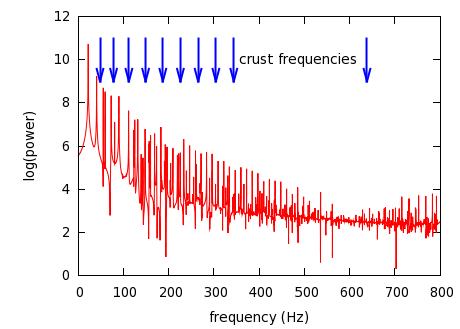}
  \end{center}
  \caption{\small Power spectrum for the dynamics of a magnetized box as described in the text. In this particular model we have used the maximum possible shear modulus, corresponding to a maximally tangled field. The Alfven motion in the box is coupled to 9 of the lowest frequency 'crustal' modes, plus a high frequency crust mode at 630 Hz.}
    \label{Fig12}
\end{figure}


We now have the basic ingredients of building a phenomenological modes with tangled
fields.
To sum up, we (1) quantify tangling using the effective shear modulus, (2) Find
discrete core eigenmodes
and evaluate their coupling to the crustal model, and (3) Either find
eigenfrequencies of the total star
by diagonalizing the potential energy of the system, or simulate the time-dependent
behavior directly.

An example of a resulting powerspectrum is shown in Fig \ref{Fig12} for the model described in this section.


\section{What do our models tell us about magnetar QPOs?}

\begin{figure}[tbp]
  \begin{center}
         \includegraphics[width=3.5in]{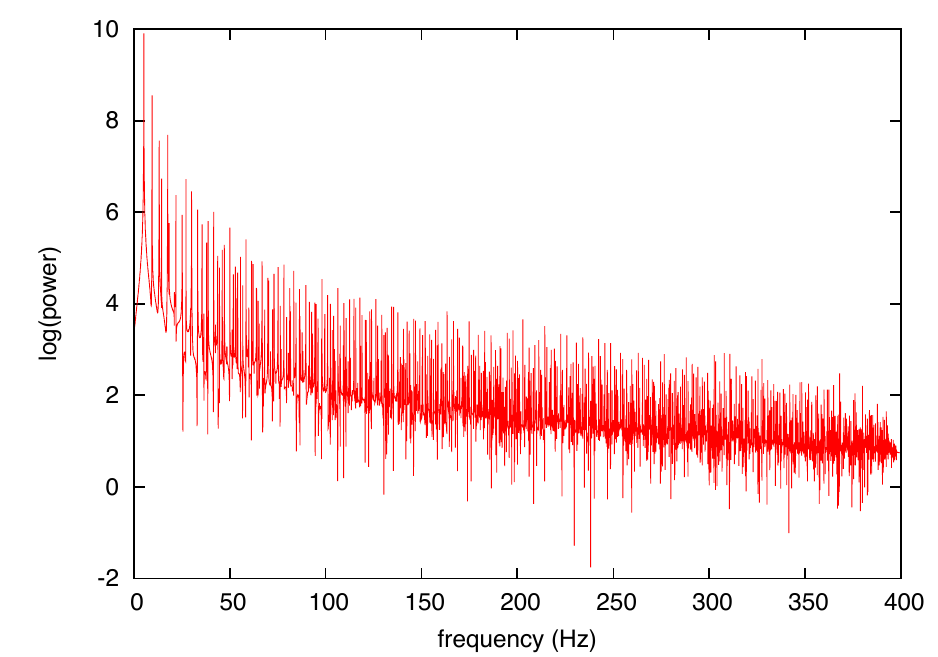}
  \end{center}
  \caption{\small This spectrum was generated using a box model similar to the one from figure \ref{Fig12} but with neutron mass-loading. Due to the mass-loading the frequencies have shifted down by a factor of $\sim 4$. Note that there is no significant power above the lower edge-mode frequency of $5.3$ Hz.}
    \label{Figspec1}
\end{figure}

\begin{figure}[tbp]
  \begin{center}
         \includegraphics[width=3.5in]{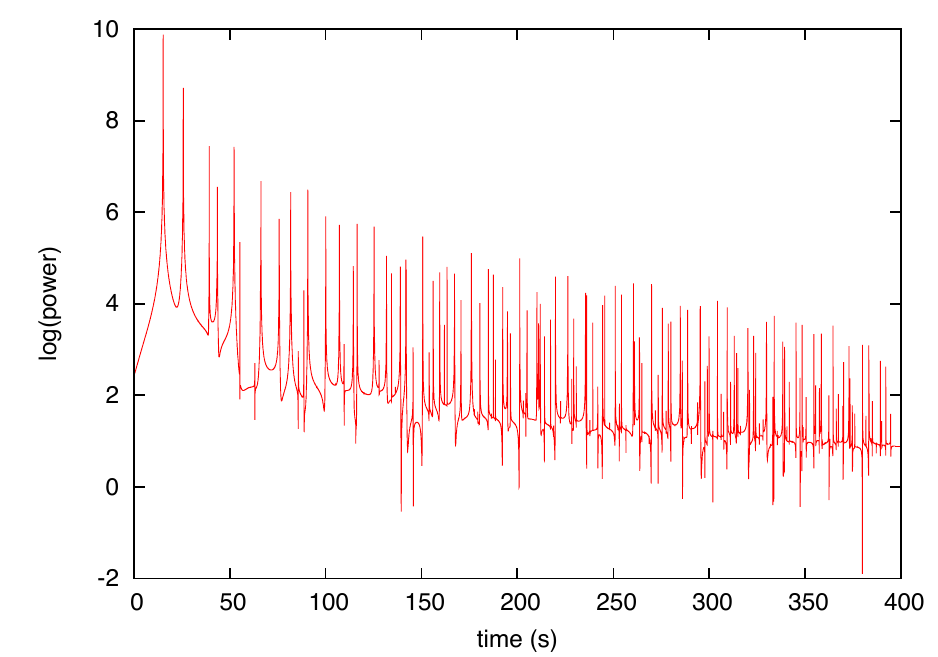}
  \end{center}
  \caption{\small This spectrum was generated with the same box model as in figures \ref{Figspec1}, but in addition to the neutron mass-loading, we have increased the magnetic field strength by a factor of 3. All frequencies above $\sim 16$ Hz are significantly damped.}
    \label{Figspec3}
\end{figure}

In this paper we have developed a formalism which allows one to build a magnetar
model with a variety
of the spectral features of the core Alfven waves, including continua with gaps and
edges, and the 
large-scale discrete
modes generated by the field tangling. We have constructed a number of magnetar models and explored the
resulting QPOS, both for the case of axisymmetric magnetar with core Alfven continuum, and for the "square" magnetar models
with the tangled fields (see the previous section). The full range of model parameters, and detailed comparison with the
data will be the subject of a separate study. For now, we have restricted ourselves to the standard magnetar model, in which the core is
a perfect conductor, the field of $\sim 10^{15}$G penetrates both the core and the crust, and the proton fraction in the star is the one
 tabulated in Haensel, Potekin, and Yakovlev (2007). 
Our models give us the following  robust conclusions, as compared against
QPO observations:

(1).  A number of strong QPOs have been observed in the 1998 and 2004 giant flares,
with frequencies ranging between 18 Hz and 150 Hz ( Israel et al.~2005, Strohmayer and Watts 2006, Watts and Strohmayer 2006). 
These QPOs can be qualitatively explained as gap and/or edge modes of sections 4 and 2, or even transient QPOs of section 
3\footnote{L07 and Gruzinov 2008b associated the long-lived 18-20Hz QPO with the lower edge of the Alfven continuum. However, 
recent calculations of Steiner and Watts (2009) have argued that the crustal frequencies can be as low as 10Hz due to the uncertainty
in our theoretical knowledge of the crustal shear modulus. It is therefore plausible that the fundamental crustal mode has the proper frequency
below that lower edge of the core Alfven continuum. In this case, the 18-20Hz QPO  could be the gap mode which is
dominated by the fundamental crustal mode.}.  However, this was only possible if the neutrons were decoupled from the Alfven waves in the core.
If the neutrons took part in the Alfven motion, then the effective mass of the Alfven modes shifted up by a factor of $20-40$
and their frequencies shifted down by a factor $4-8$ (Easson \& Pethick 1979, Alpar et al.~1984,
van Hoven \& Levin 2008, Andersson et al.~2009). As a result, all modes at frequencies above $\sim 50$Hz were strongly damped
(see Fig. \ref{Figspec1}). Increasing the magnetic-field tension by a factor of 3 did not affect this conclusion (Fig \ref{Figspec3}). For the spherical magnetar models of section 4 we obtain similar results if couple the neutrons to the Alfven motion in the core.
The key point that we would like the reader to appreciate is  that Alfven modes in the core are key to determining the frequency and strength of the  
observable QPOs, and thus QPOs are very sensitive probe of the core interior.

(2) A number of the high-frequency QPOs have been measured in the 2004 giant  flare by Watts and Strohmayer (2006), the strongest among them
being the QPO at $625$ Hz. This QPO is particularly strong and long-lived in the hard x-rays, reaching the amplitude of $\sim 25$\% over the time
interval of $\sim 100$ seconds (i.e., it persists for almost $10^5$ oscillation periods!).  Watts and Strohmayer (2006) argued that 
this frequency corresponds to the crustal shear mode with a single radial node (see also Piro 2005); this interpretation, if correct, would
strongly constrain the thickness of the crust and rule out the fluid strange stars as magnetar candidates (Watts \& Reddy, 2007).
To investigate this suggestion, we have introduced several high-frequency low-j crustal modes into our square-box 
simulations. However, as is demonstrated in Figs. \ref{Fig9} and \ref{Fig10}, the high-frequency modes are strongly damped and at no time during the simulations
do we observe any significant power at those frequencies. This is to be expected. No natural axisymmetric model has gaps in the Alfven continuum 
at such high frequencies, so global modes are strongly absorbed. One could expect that if the Alfven modes are discrete in the core due to the field tangling, 
this problem would not arise. However, even in the discrete case the frequency spacing between the modes is around $20$ Hz, which is much smaller
than $600$ Hz. Thus the grid of Alfven waves is so dense that it is effectively seen as the absorbing continuum by the modes around  $600$ Hz. Our detailed
simulations, of the type shown in Figs. \ref{Fig9} and \ref{Fig10},
fully confirm this qualitative picture.

The concern about the viability of high-frequency QPOs as being due to the physical oscillations of standard-model magnetars 
has been raised in the original L06 paper
on the basis of rather simplistic calculations. As  our work here shows, more detailed calculations partially alleviate the L06 concern for the frequencies
below $\sim 150$Hz, but only if the neutrons are decoupled from the Alfven motion in the core, i.e. if at least one baryonic superfluid (protons or neutrons)
are present in the neutron-star core. Our analysis sustains L06 concern for the high-frequency QPOs, in particular for the strong long-lived QPO at 625Hz.
Its explanation seems to require either QPO production in the magnetosphere, or a somewhat radical revision of the magnetar model. Just how radical
this revision has to be will be explored in a separate study.

Our work presented here has several shortcomings.
We have  limited  ourselves to the linear approximation, 
and a non-linear regime may bring surprises. Direct non-linear simulations
of axisymmetric  oscillations of a magnetised fluid star has been carried out recently by  Cerda-Duran, Stergioulas, \& Font (2009). At this stage it is difficult
to say whether non-linearities introduce significantly  new QPO features to their model; their results have largely been in agreement with the linear
simulations of Colaiuda, Beyer, \& Kokkotas (2009). However, the computational techniques seem promising and we do not exclude that
large-amplitude simulations of
stars with the crust will show qualitatively new features. Another limitation of our work is that we have assumed that once the flare sets the magnetar
into motion, the magnetar's oscillations are not driven externally. This may not be the case in real flares: some energy stored in the pre-flare magnetar
may be released gradually, and this release could be extended in time into the flare's tail\footnote{We thank Chris Thompson for pointing out this possibility.}.
The latter consideration is straightforward to incorporate phenomenologically into our model, and we plan to address it in our future work.  

\section{Acknowledgements}
We thank Andrei Beloborodov, Anna Watts, 
Peter Goldreich, Chris Thompson, and especially Andrei Gruzinov for useful discussions. This research has been supported by
Leiden Observatory and Lorentz Institute for theoretical physics through internal grants.

\section*{References}
\begin{footnotesize} \noindent
Alpar, M.~A., Langer, S.~A., \& Sauls, J.~A.~1984, ApJ, 282, 533\\
Andersson, N., Glampedakis, K., \& Samuelsson, L.~2009, MNRAS, 396, 894\\
Barat C. et al., 1983, A\& A, 126, 400\\
Braithwaite, J.~2008, MNRAS, 386, 1947\\
Braithwaite, J., \& Spruit, H.~C.~2004, Nature, 431, 819\\
Braithwaite, J., \& Nordlund, A.~2006, A\&A, 450, 1077\\ 
Cerda-Duran, P., Stergioulas, N., \& Font, J.~2009, MNRAS, 397, 1607\\
Colaiuda, A., Beyer, H., \& Kokkotas, K.D.~2009, MNRAS, 396, 1441\\
Duncan, R.~C.~1998, ApJ, 498, L45\\
Easson, I., \& Pethick, C.~J.~1979, 227, 995\\
Glampedakis K., Samuelsson L.,  \& Andersson N., 2006, MNRAS, 371, L74\\
Goedbloed, J.~P.~H., \& Poedts, S.~2004, Principles of Megnetohydrodynamics 
(GP in the text; Cambridge University Press)\\
Gruzinov, A., 2008a, arXiv: 0801.4032 [astro-ph]\\
Gruzinov, A., 2008b, arXiv: 0812.1570 [astro-ph]\\
Gruzinov, A., 2009, arXiv: 0905.0911 [astro-ph]\\
Haensel P., \& Potekhin A.~Y., 2004, A\&A, 428, 191\\
http://www.ioffe.ru/astro/NSG/NSEOS/\\
Haensel P., Potekhin A.~Y.,  \& Yakovlev D.~G., 2007. Neutron Stars 1: Equation of State
and Structure (New York: Springer)\\
Hollweg, J.~V.~1987, ApJ, 312, 880\\
Israel G.~L. et al., 2005, ApJ, 628, L53\\
Ionson, J.~A.~1978, ApJ, 226, 650\\
Landau, L.~D., \& Lifshits, E.~M.~1976, Mechanics, chapter V
 (Pergamon press)\\
Lee, U.~2008, MNRAS, 385, 2069\\
Levin Y., 2006, MNRAS Letters, 368, 35 (L06 in the text)\\
Levin Y., 2007, MNRAS, 377, 159 (L07 in the text)\\
Piro, A.~2005, ApJ, 634, L153\\
Poedts, S., Hermans, D., \& Goossens, M.~1985, A\&A, 151, 16 (P85 in the text)\\
Samuelsson, L., \& Andersson, N.~2007, 374, 256\\
Sotani, H., Kokkotas, K.~D., \& Stergioulas, N.~2008, MNRAS Letters, 385, 5\\
Steiner W., \& Watts A.~L., 2009, Phys.~Rev.~Letters, 103r1101S\\
Strohmayer, T.~E. \& Watts, A.~L., 2005, ApJ, 632, L111\\
van Hoven, M., \& Levin, Y.~2008, MNRAS, 391, 283\\
Watts, A.~L. \& Strohmayer T.~E., 2006, ApJ, 637, L117\\
Watts, A.~L., \& Reddy, S.~2007, MNRAS Letters, 379, 63\\

\end{footnotesize}


\appendix

\section{Multimodal crust-core system}
In this Appendix we generalize the normal-mode treatment of Section 2.2,
and write down the general prescription of how to find the
eigenmodes when {\it several} ``large'' crustal shear modes are coupled to a multitude
of small core Alfven modes, provided the coupling
coefficients are known. In this paper, the coupling 
coefficients are worked out in simple models of sections 4 and 5; 
we shall postpone the discussion of
how the coefficients are computed in more general case to the future paper. 

Let us denote the displacement of the crustal and core modes by $X_n$ and $x_i$ respectively.
Since both the crustal and the core modes are not directly coupled to themselves (i.e.,
$X$'s are only coupled to $x$'s),  most general equations of motion take the form

\begin{eqnarray}
\ddot{X}_n+\Omega^2 X_n&=&\Sigma_i \alpha_{ni}x_i\label{multimode1}\\
\ddot{x}_j+\omega_j^2 x_i&=&\Sigma_m \beta_{jm}X_m, \nonumber
\end{eqnarray} 
where $\Omega_n$ and $\omega_j$ are the proper frequencies of the crustal and
core modes, and $\alpha$'s and $\beta$'s are the coupling coefficients.
We look for an oscillatory solutions of the above equations with angular frequency
$\Omega$. One can trivially re-write these equation as a matrix eigen-equation
with $\Omega^2$ as an eigenvalue, and solve it using standard methods. However,
if the number of crustal modes is not too large, it is convenient to make a shortcut.
Using the second of Eq.~(\ref{multimode1}) to express $x_i$'s through $X_n$'s, and
substituting into the first one,
we get the following equation:
\begin{equation}
\Sigma_n G_{mn}(\Omega)X_n=0,
\end{equation}
where the elements of the matrix $G$ are given by
\begin{equation}
G_{mn}(\Omega)=(\Omega^2-\Omega_n^2)\delta_{nm}+\Sigma_i{\alpha_{ni}\beta_{im}\over \omega_i^2-\Omega^2}.
\end{equation}
One obtains the eigenfrequencies by finding numerically the zeros of $\det {G_{mn}}$.

\section{Core continua with a mixed axisymmetric toroidal-poloidal magnetic field}
In this appendix we will calculate the continuum of Alfven frequencies in a magnetar core in the case of a axisymmetric magnetic field with mixed toroidal and poloidal components. The general MHD equations of motion for spherically symmetric, self-gravitating equilibrium with an axisymmetric field, are derived in detail in P85. In contrast to the special case of a purely poloidal field (see section 4.2) which leads to two uncoupled differential equations, the continuum for a mixed toroidal-poloidal field is described by a system of fourth order coupled ODEs. Due to this coupling, the solutions are complicated as they are no longer polarized in the directions parallel (so-called "cusp solutions") and perpendicular (Alfven solutions) to the magnetic field lines, but rather have a mixed character. Strictly speaking, one can only speak of an "Alfven continuum" in the limit that the variations in $\rho$, $P$ and $B^2$ are small in the magnetic flux-surfaces. The general equations of motion are given in Eqs. (53) and (54) of P85. We note however, that in magnetars the speed of sound $c >> c_A$, and therefore we consider Poedts et al.'s equations (53) and (54) in the incompressible limit (P85, Eqs. (73) and (74)). For completeness we give the equations here, 

\begin{equation}
\rho \sigma^2 \frac{B_{\chi}^2 B^2}{B_{\phi}^2} Y = B^2 F\frac{B_{\chi}^2}{B_{\phi}^2 B^2} F \left( \rho c_A^2 Y \right) + \frac{1}{\rho c_A^2} \left[ \frac{\partial}{\partial \chi} \left( \rho c_A^2 \right) \right]^2 Y +\rho B_{\chi}^2 N_{\chi}^2 \left( Y+Z \right) - \frac{\partial}{\partial \chi} \left( \rho c_A^2 \right) F Z
\label{B1}
\end{equation}

\begin{equation}
\rho \sigma^2 B^2 Z = i F \left[ \frac{\partial}{\partial \chi} \left( \rho c_A^2 \right) Y \right] + \rho B_{\chi}^2 N_{\chi}^2 \left( Y+Z \right) + F \left( \rho c_A^2 FZ \right)
\label{B2}
\end{equation}

The variables $Y \equiv i \left(B_{\phi}^2 \xi_{\chi} - B_{\phi}B_{\chi}\xi_{\phi}\right)/B_{\chi}B^2$ and $Z \equiv i \left( B_{\chi}\xi_{\chi} + B_{\phi}\xi_{\phi} \right)/B^2$ are components of the fluid displacement perpendicular and parallel to the magnetic field lines, the operator $F \equiv i \partial / \partial \chi$ is a differential operator along the field lines, $N_{\chi} \equiv -\left( 1/B_{\chi}\rho \right) \sqrt{\left(\partial \rho /\partial \chi \right) \left(\partial P /\partial \chi \right)}$ can be thought of as a Brunt-V\"ais\"al\"a frequency for displacements along the field lines. According to Gauss' law for magnetism, the toroidal component of the magnetic field is of the form $B_{\phi} = f(\psi )/ \varpi$, where $\varpi$ is the distance to the polar axis, and $f(\psi )$ is an arbitrary function of $\psi$. In the following calculation we adopt a toroidal field component of the form 

\begin{equation}
B_{\phi} = \frac{B_{\rm t, 0} R_*}{\varpi \left( \chi \right)}  \sin{\left(\theta (\psi)\right)},
\label{Btor}
\end{equation} 

Here $\theta (\psi)$ is the polar angle at which the flux surface $\psi$ intersects the stellar crust. Clearly this choice for $B_{\phi}$ is completely arbitrary and one could in principle try many different toroidal geometries. \\

\begin{figure}[tbp]
\centering
       \includegraphics[width=4in]{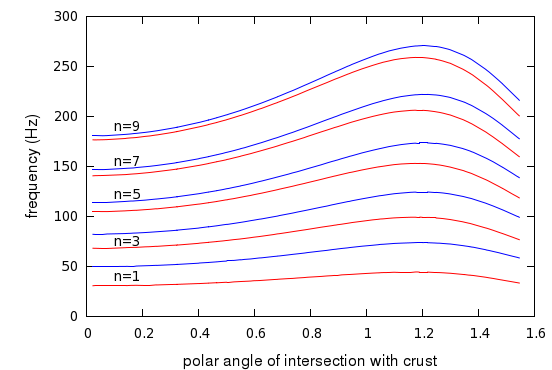}
  \caption{\small The curves show the continuum frequencies $\sigma_n$ as a function of the angle $\theta (\psi)$, the polar angle at which the flux-surface $\psi$ intersects the crust. In the presence of a toroidal field, the degeneracy between the cusp-solutions and the Alfven solutions is broken and we find two 
  separate solutions for each wave number $n$; waves with primarily Alfven character (red curves) and waves with primarily cusp character (blue curves). 
  This particular continuum was calculated using a poloidal field with an average surface value $B_{\rm p, surface} \sim 6 \cdotp 10^{14}~ \rm{G}$ 
  (again generated by a circular ring current of radius $a = R_{*}/2$) and a toroidal field strength at the equator and the crust-core interface of 
  $B_{\rm t, 0} = 3 \cdotp 10^{14}$ G (see Eq.~(\ref{Btor})).}
    \label{pt_freqs}
\end{figure}

As with our calculation of the Alfven continuum in the case of a purely poloidal field (section 4.2), we adopt the zero-displacement boundary conditions at the crust, and use the fact that our equilibrium model is (point-) symmetric with respect to the equatorial plane. This enforces the existence of classes of symmetric and anti-symmetric eigenfunctions, $Y_n(\chi)$ and $Z_n(\chi)$. We consider only the odd modes, and calculate the eigenfunctions by means of the shooting method; we use a fourth order Runge-Kutta scheme to integrate Eqs. (\ref{B1}) and (\ref{B2}). Starting with $Y(0) = 0$ and $Z(0) = 0$ at the equator, we integrate outward until we reach the crust at $\chi = \chi_c$. We find the eigenfrequencies by changing the value of $\sigma$ until we match the boundary conditions at the crust. A resulting continuum is plotted in Figure \ref{pt_freqs}.

\clearpage

\end{document}